\documentclass[10pt,conference]{IEEEtran}
\usepackage{cite}
\usepackage{amsmath,amssymb,amsfonts}
\usepackage{textcomp}
\usepackage{algorithmicx}
\usepackage{algcompatible}

\usepackage{graphicx}
\usepackage{textcomp}
\usepackage{xcolor}
\usepackage[hyphens]{url}
\usepackage{fancyhdr}
\usepackage{hyperref}
\usepackage{soul}
\usepackage{comment}
\usepackage[noend]{algpseudocode}
\usepackage{algorithm}
\usepackage{algorithmicx}
\newcommand{\hpcayear}{2024}

\newcommand{\hpcasubmissionnumber}{NaN}
\title{A detailed algorithmic study on a reuse-aware, near memory,  all-digital Ising machine}

\newcommand\hpcaauthors{Siddhartha Raman Sundara Raman, Lizy K. John, Jaydeep P. Kulkarni}
\newcommand\hpcaaffiliation{The University of Texas at Austin}
\newcommand\hpcaemail{s.siddhartharaman@utexas.edu}

\author{
  \ifdefined\hpcacameraready
    \IEEEauthorblockN{\hpcaauthors{}}
      \IEEEauthorblockA{
        \hpcaaffiliation{} \\
        \hpcaemail{}
      }
  \else
    \IEEEauthorblockN{\normalsize{HPCA \hpcayear{} Submission
      \textbf{\#\hpcasubmissionnumber{}}} \\
      \IEEEauthorblockA{
        Confidential Draft \\
        Do NOT Distribute!!
      }
    }
  \fi 
}

\fancypagestyle{camerareadyfirstpage}{%
  \fancyhead{}
  
  \fancyhead[C]{
    \ifdefined\aeopen
    \parbox[][12mm][t]{13.5cm}{\hpcayear{} IEEE International Symposium on High-Performance Computer Architecture (HPCA)}    
    \else
      \ifdefined\aereviewed
      \parbox[][12mm][t]{13.5cm}{\hpcayear{} IEEE International Symposium on High-Performance Computer Architecture (HPCA)}
      \else
      \ifdefined\aereproduced
      \parbox[][12mm][t]{13.5cm}{\hpcayear{} IEEE International Symposium on High-Performance Computer Architecture (HPCA)}
      \else
      \parbox[][0mm][t]{13.5cm}{\hpcayear{} IEEE International Symposium on High-Performance Computer Architecture (HPCA)}
    \fi 
    \fi 
    \fi 
    \ifdefined\aeopen 
      \includegraphics[width=12mm,height=12mm]{ae-badges/open-research-objects.pdf}
    \fi 
    \ifdefined\aereviewed
      \includegraphics[width=12mm,height=12mm]{ae-badges/research-objects-reviewed.pdf}
    \fi 
    \ifdefined\aereproduced
      \includegraphics[width=12mm,height=12mm]{ae-badges/results-reproduced.pdf}
    \fi
  }
  \fancyfoot[C]{}
}
\begin{document}
\def\hpcacameraready{1}
\maketitle

  \thispagestyle{plain}
  \pagestyle{plain}

\newcommand{\hpcaheight}{0mm}
\ifdefined\eaopen
\renewcommand{\hpcaheight}{12mm}
\fi


\begin{abstract}

Recently, nature-inspired computing approaches have gained significant attention for solving difficult optimization problems, particularly through Ising machines for NP-complete applications. Existing Ising accelerators range from quantum and optical annealers to CMOS-based von-Neumann and in-memory architectures. However, many prior designs are specialized accelerators limited to specific problem classes, rely on ADC/DAC circuits, and suffer from reliability challenges due to process-variation-sensitive embedded memory technologies.

This paper presents SACHI, an all-digital Ising architecture implemented by repurposing the L1 cache of a CPU using SRAM-based processing-in-memory techniques. SACHI eliminates the need for ADCs/DACs, improves reliability compared to prior approaches such as BRIM, and enables Ising acceleration with minimal hardware overhead integrated into the CPU pipeline. The paper also provides detailed architectural analysis and pseudo-code for the proposed algorithms.

The key contributions of SACHI are: (i) tight integration of the accelerator with the CPU pipeline, (ii) reuse of existing cache hardware for acceleration, (iii) higher parallelism enabled through reuse-aware computation, and (iv) improved performance and energy efficiency for large-scale, high-precision optimization problems using novel compute and mapping strategies.

Compared to BRIM, SACHI achieves 300x performance improvement and 80x energy reduction across applications including asset allocation, molecular dynamics, image segmentation, and traveling salesman problems. Additionally, reuse factors up to 4000x are observed for several workloads. This work demonstrates that reliable and efficient all-digital Ising acceleration can be achieved using commodity SRAM structures tightly integrated with general-purpose processors.  

\end{abstract}

\section{Introduction}
\begin{figure}[t]
\centering
\includegraphics[width=\linewidth]{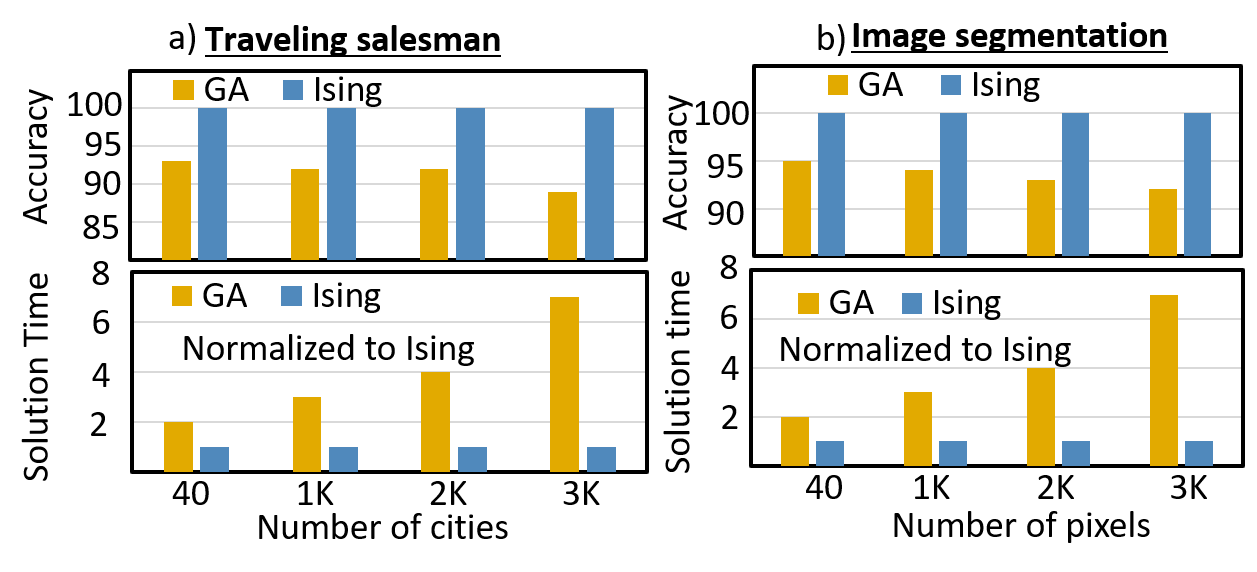}
\vspace{-2em}
\caption{\textbf{{Ising machines provide higher accuracy and lower solution time. (i) Top - Solution accuracy under iso-performance condition, and (ii) Bottom - execution time normalized to Ising under iso-accuracy, comparing  Genetic Algorithm (GA) and Ising for a) traveling salesman, and b) image segmentation problems.}}}
\label{fig:Motivation}
\vspace{-1em}
\end{figure}

The quest for efficient solutions to complex combinatorial optimization problems (COP) has been an enduring pursuit in the realm of computing and scientific research.  Classical algorithms like brute-force search or gradient-based methods, widely employed for optimization, struggle to efficiently explore the vast solution spaces and find optimal solutions within a reasonable timeframe for real-life applications. In this context, iterative approaches like genetic algorithm, particle-swarm optimization, etc have emerged as front-runners, displaying ability to achieve faster convergence towards optimal solutions. 
Recently, there has been a surge in research focused on solving combinatorial optimization problems (COPs) by drawing inspiration from nature's principles or harnessing natural phenomena. A significant illustration of such an approach is the utilization of Ising machines
\cite{isinghistory}\cite{timedepstats}, which hold great promise in tackling NP-complete optimization problems. Ising machines~\cite{BRIM, ISCA_2022}  have emerged as a promising frontier, offering innovative approaches that leverage the principles of statistical mechanics to represent and solve these optimization problems efficiently. These are well-suited for optimization problems \cite{Ising_formulation}, particularly those involving minimizing energy in physical systems, such as max-cut, asset allocation, graph partitioning, traveling salesman, etc. These differ from classical machines, as classical machines are versatile and can handle a wide range of computational tasks, from scientific simulations/data analysis to general-purpose computing. Ising machines use concepts of spins and interaction coefficients (IC) to represent and solve optimization problems efficiently \cite{ferro}. They encode variables and constants as spins and ICs, respectively. They utilize the Hamiltonian energy function as a heuristic to find optimal solutions to COPs. For instance, Ising machines can achieve more accurate solutions in less time compared to other heuristic-based optimization algorithms, like genetic algorithms. 
Fig.\ref{fig:Motivation} illustrates the solution accuracy of genetic algorithms (GA) \cite{GA} and Ising machines for traveling salesman and image segmentation problems in the top two figures. Ising machines are seen to provide greater than 99\% accuracy whereas genetic algorithms achieve less than 95\% accuracy. In iso-accuracy scenarios, the solution time for Ising is 2x-6x smaller than GA. 

\par From an architectural standpoint, the distinctions in the representation of spins and ICs give rise to different Ising machine implementations, which have been realized using physical \cite{ISCA_2022} or iterative models~\cite{Shanshan_Ising}.
The traditional realization of Ising machines captures the dynamics of the physical Ising model by using qubits\cite{Quantum_annealing}\cite{experimental_coherent}, coupled oscillators\cite{Oscillator_Ising}\cite{wang2019oim}, and optical annealers\cite{Optical_Ising}\cite{optical_time}. An excellent summary of three and a half generations of Ising machines is provided in ~\cite{ISCA_2022}; hence we avoid a detailed description here. The major challenge of these approaches is the need for cryogenic operating temperature in qubits, increased power requirement in coupled oscillators, and high area requirement in optical annealers. Another approach involves the usage of CMOS-based Von-Neumann-like iterative Ising machines\cite{Hitachi}, which perform iterative updates to the spins to achieve the approximate ground state solution. However, the issue is that in a real-life application consisting of many variables, extensive energy is spent on the data movement of variables. In order to reduce the data movement costs, computing in/near memory (CIM/CNM) based Ising machines are being investigated. The advantage of CIM designs is that modern high-density, inexpensive on-chip memory is repurposed to map large-sized COP onto them for performing efficient in-memory compute. 
 \par The prevailing challenges with the existing state-of-the-art accelerators, whether in the physical approach (BRIM)~\cite{BRIM} or the iterative approach (Ising-CIM)~\cite{Shanshan_Ising}, are multi-faceted.
These are dedicated domain-specific accelerators optimized for only a specific subset of COPs, and do not support different resolutions for efficient compute. Furthermore, their reliance on analog data converters or blocks render them susceptible to process variations, causing reliability issues. In addition, the lack of reuse in them leads to increased data movement, further exacerbating energy/performance concerns.

\begin{figure}[t]
\centering
\includegraphics[width=\linewidth]{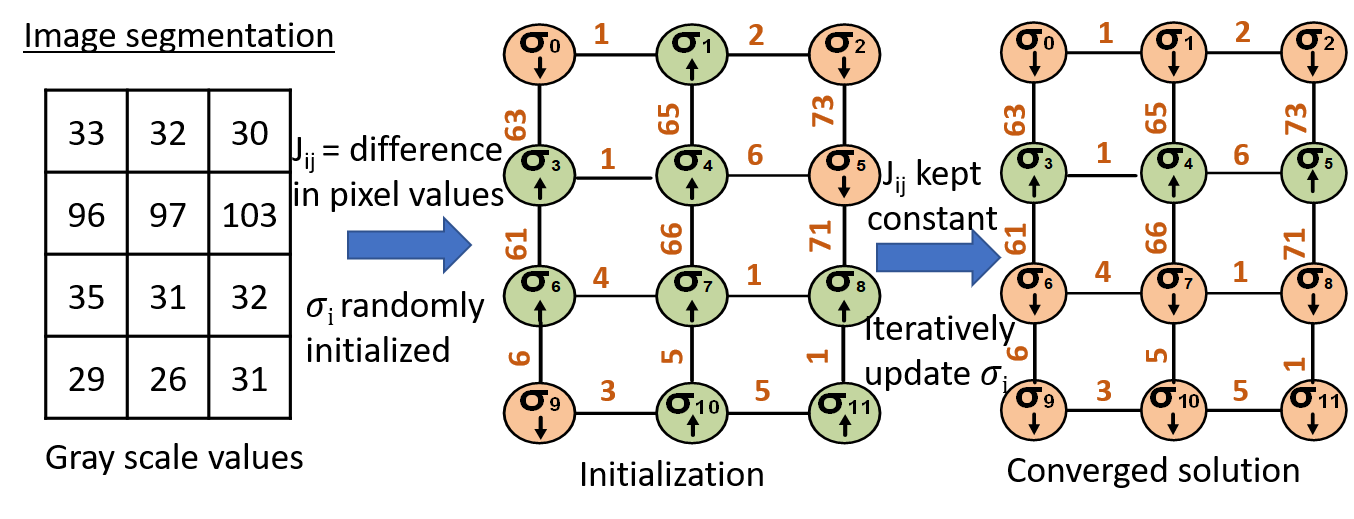}
\vspace{-2em}
\caption{\textbf{\underline{COP workload mapping to Ising Machine} - Image segmentation example for 4*3 image with the edges showing interaction coefficient as a difference between pixel values. Spin of +1 indicated by green) and -1 by orange.
Spins are randomly initialized and Ising machine converges to a segmented image. }} 
\label{Graph_type}
\vspace{-1em}
\end{figure}

The major contributions of the paper are: 

\begin{itemize}
\item{Architecting an all-digital Ising machine that repurposes the L1 cache hardware for in-memory compute, and presenting an accelerator that is tightly coupled to the CPU pipeline, thereby minimizing extra hardware}
\item{A reuse-aware computing strategy along with multiple data-stationary PIM designs that leverage the compute strategy to perform less redundant compute, achieving high parallelism and energy efficiency}
\item {A tuple mapping strategy is proposed to abstract the incoming graph structure, that makes SACHI scalable to any graph used to represent large-size real-life COPs}
\item {A mixed encoding scheme to enable SACHI to be reconfigurable to any precision upto 32-bit for in/near-memory compute, without the usage of DACs/ADCs}


\item{Evaluation of SACHI using several real-world complex COPs indicate that  SACHI offers 160x/36x/286x/300x better performance and 79x/72x/80x/75x better energy over BRIM  for molecular dynamics/asset allocation/image segmentation/traveling salesman problems. Furthermore, SACHI offers a speedup of 90x and energy improvement of 75x over Ising-CIM. } 

\end{itemize}

\begin{figure}[t]
\centering
\includegraphics[width=8.9cm]{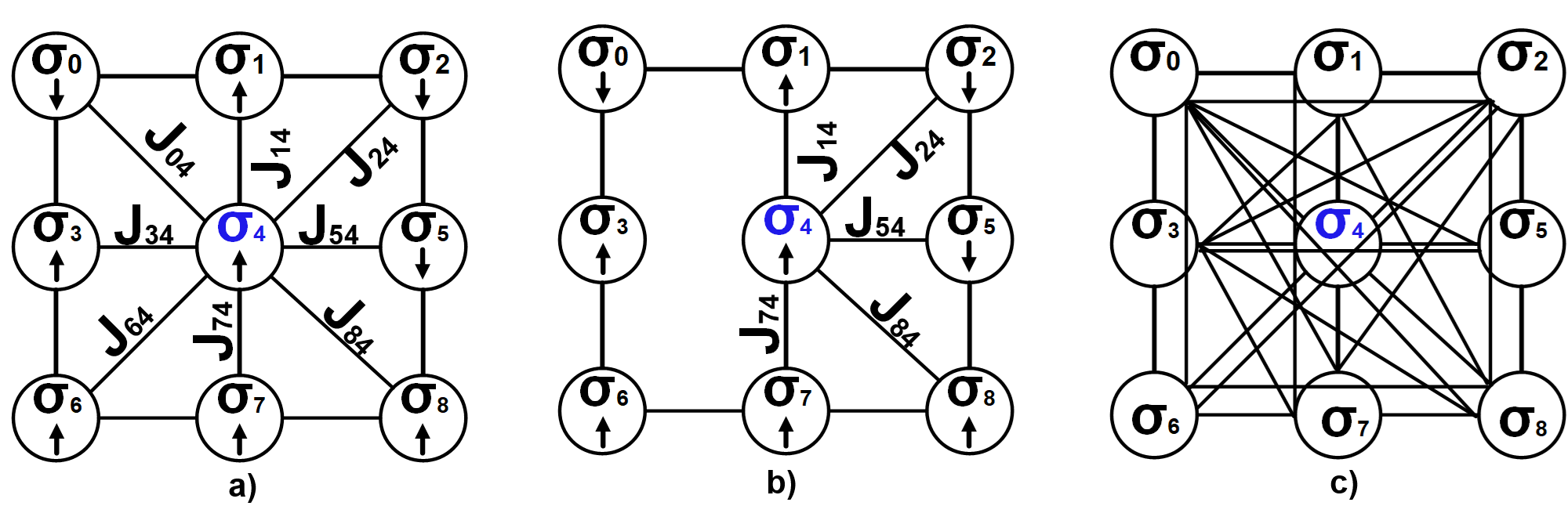}
\vspace{-2em}
\caption{\textbf{a) 3*3 King's Graph, commonly used for mapping Ising models b) Sparsely connected 3*3 Graph c) Fully connected 3*3 Graph. The target spin is shown in blue in all the figures}}
\label{Graph_network}
\vspace{-1em}
\end{figure}
\subsection{Mapping COP onto Ising Model} 
The Ising model\cite{ising_machine}\cite{ising_model_intro}, originating from statistical mechanics, is used to study the alignment of spin orientations (either up-spin or down-spin) in a magnetic material under the presence of external perturbations. Individual spins interact with one another and flip their orientations, so that the collection of spins in a magnetic material reach a minimum ensemble energy state (called the `ground state') \cite{Shanshan_Ising}. The spins and ICs represent the variables and the relationship between these variables, respectively, in a COP. The minimum ensemble energy state represents the optimal solution to COP. For image segmentation\cite{image_seg}, IC identifies the edge value between 2 neighboring pixels (spins) by finding the difference between them, with the mapping onto Ising model (see Fig.\ref{Graph_type}).
For traveling salesman \cite{TRS}, IC represents the distance between the 2 cities (spins). For asset allocation, which investigates the feasibility of splitting a net worth of USD X Million valued across N assets (spins) among people, IC is the value of each asset allocated. The optimal solution to COPs for iterative Ising model is obtained by minimizing Hamiltonian energy\cite{hamiltonian}, a function of pair-wise coupling among those spins, shown in Fig.\ref{Graph_network} given as: 
\begin{equation}
\vspace{-1em}
  H = -\sum_{ij}^{} J_\mathrm{ij} \sigma_\mathrm{i} \sigma_\mathrm{j} - \sum_{i}^{} h_\mathrm{i} \sigma_\mathrm{i} 
\vspace{1em}
\end{equation}
\par where J\textsubscript{ij} represents the ICs, $\sigma_\mathrm{i}$ represents the target spin (for which the update is being performed), $\sigma_\textsubscript{j}$ represents the neighboring spins of $\sigma_\mathrm{i}$, h\textsubscript{i} represents the external field, with i, j representing a pair of nodes in a graph. 
The minimization of H carried out by a divide and conquer update of each spin, based on its interaction with its neighbors \cite{CIM-Spin} results in:
\begin{equation} \label{eq:h_sigma}
    H_\mathrm{\sigma} = \sum_{}^{} -J_\mathrm{ij} * \sigma_\mathrm{j} - h_\mathrm{i} 
\end{equation}
The \textbf{spin update} is carried out based on the sign of H\textsubscript{$\sigma$}: 
\vspace{-0.5\baselineskip}
\begin{align}\label{eq:spin_update_sigma}
   \sigma = 
\begin{cases}
    -1,     & \text{if } H_\mathrm{\sigma} > 0\\
    +1,     & H_\mathrm{\sigma} < 0 \\
    +1/-1   & H_\mathrm{\sigma} = 0   
\end{cases}
\end{align}
\begin{figure}[t]
\centering
\includegraphics[width=\linewidth]{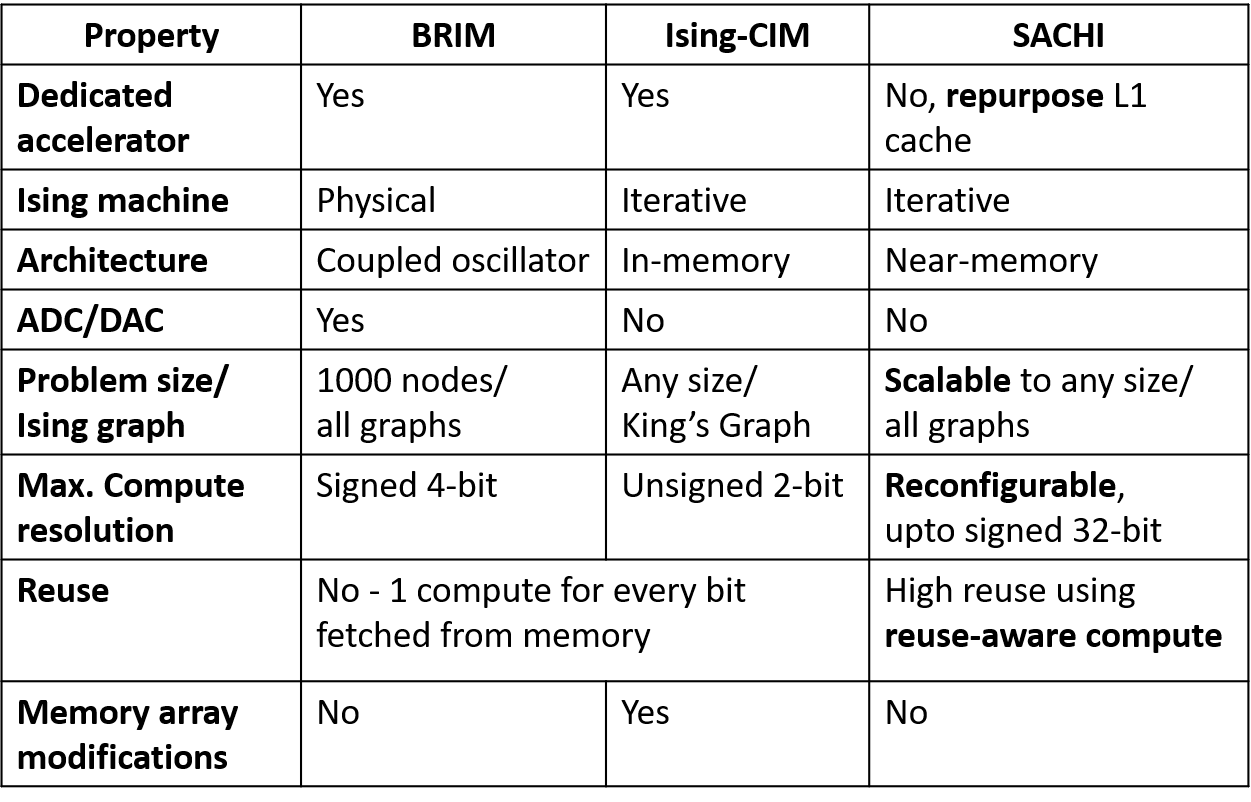}
\vspace{-2em}
\caption{\textbf{{SACHI in comparison to to state of the art Ising architectures BRIM and Ising-CIM} - SACHI is a repurposable, scalable, reconfigurable near-memory architecture with no modifications to the memory array, no ADCs/DACs, achieving better reuse/energy as compared to prior Ising accelerators like BRIM\cite{BRIM}, Ising-CIM \cite{Shanshan_Ising} }}
\label{Graph_network}
\vspace{-1em}
\end{figure}

\par The local spin update might result in H being trapped in a local minimum. Simulated annealing is then performed to achieve the global minima by probabilistic spin-flips.

\subsection{Simulated annealing}
\par The algorithm starts with exciting the spins in the model to a high energy state and then decreasing the energy slowly, thereby leading to an optimal solution. One approach is the usage of pseudo- temperature to identify the likelihood of spin update, given by equation \ref{eq:spin_update_sigma}. The pseudo temperature is reduced towards zero from an initial temperature(initT), with several iterations to progressively decrease the number of spin flips. The likelihood function typically makes use of the objective function to be minimized, as shown in alg.1. Here, the objective function is the Hamiltonian energy of the system. The metropolis acceptance criterion  for accepting the spin update value is used, i.e. the likelihood is compared against a randomly generated value, to decide on the spin value post update. These probabilistic spin flips are indispensable for avoiding H being trapped in local minima. The overall algorithm for minimization of energy of spin is shown in Algorithm 1.

\begin{algorithm}[b]
\caption{PseudoCode for minimization of energy of spin}
\begin{algorithmic}[1]

\Procedure{Spin Update}{$G_s$}
\Comment{$G_s$ contains $s_i, s_j, J_{ij}$}

\For{$\langle s_n, j_n \rangle \in \langle s_j, J_{ij} \rangle$}

    \State $H_{\sigma} \gets H_{\sigma} + s_n \times j_n$
    \State $s \gets 1$

    \If{$H_{\sigma} \geq 0$}
        \State $s \gets -1$
    \EndIf

\EndFor

\State \Return $s$
\EndProcedure

\Procedure{Sim Anneal}{$iterNum, updS, currS, initT, l$}

\State $Temperature \gets \frac{initT}{iterNum}$

\State $likelihood \gets
\exp\left(-\frac{H(updS)-H(currS)}{Temperature}\right)$

\If{$l < likelihood$}
    \State $currS \gets updS$
\Else
    \State $currS \gets -updS$
\EndIf

\EndProcedure

\Procedure{Min Energy}{$G_s, iterNum, currS, initT, l$}

\State $updS \gets \Call{Spin Update}{G_s}$

\State \Call{Sim Anneal}
{$iterNum, updS, currS, initT, l$}

\EndProcedure

\end{algorithmic}
\end{algorithm}
\subsection{Related work}
Ising machines have been realized either by using physical Ising models or minimizing Hamiltonian energy by using an iterative spin update method. Physical Ising models are modelled using 1) Quantum annealers, marketed by D-wave \cite{D-wave}, 2) Optical Ising annealers \cite{Optical_Ising} 3) CMOS based coupled oscillator models \cite{Oscillator_Ising}\cite{Ising_summary}. Quantum annealers use qubits to encode information, and the ground state of the system is identified by the collective state of qubits. However, these are susceptible to noise and hence require operation at cryogenic temperatures. The need for devices capable of operating at such low temperature results in additional cooling cost. Optical Ising annealers make use of a phase modulator to encode binary spin states, and the ground state is determined by the amount of propagation of light. However, these are limited by the physical size of these devices and hence, face scaling issues as the complexity of the models increase.  On the other hand, CMOS based coupled oscillator models overcome the disadvantages of 1) and 2) as they can be operated at room temperature and use smaller sized components. In this case, the ground state is determined by the phase of coupled oscillators. However, there is a surge in the total power, predominantly arising from high toggling rate at the oscillator nodes hindering the scalability of CMOS based oscillators to complex graph networks. In the resistive coupled oscillator approach (BRIM), spins are stored in capacitors, and resistances are programmed according to ICs (disadvantages in Sec.III). We would like to mention that the philosophy of physical Ising machine is completely different from that of iterative Ising machines. The main reason for comparing with BRIM is that it is the only prior work in this domain familiar to architects.  
\par Iterative Ising models are widely being researched for their simplicity in mapping the CO problems to CMOS based annealers. The first generation Ising machine, proposed by Hitachi\cite{Hitachi} used a Von-Neumann style of architecture with readout of spins and interaction coefficients from memory during every cycle, followed by computation using dedicated arithmetic logic gates located far from memory. This approach suffers from higher data movement, minimal scalability and lower energy efficiency. In order to reduce data movement, digital annealers were presented using Compute in memory approach in \cite{CIM-Spin}. In \cite{CIM-Spin}, the building block is a compute in memory(CIM) based column structure and multiple column structures are integrated together to form the CMOS Ising chip. Each column structure consists of 4 SRAM bitcells and CIM logic. The CIM logic consists of dedicated logic circuits for 1) computing partial dot product using XOR gates, 2) adders for accumulating the partial dot products. Developing the CIM structure involves modifications to the memory array and degrading the storage density (usually measured in Mb/mm\textsuperscript{2}. Also, the accumulation of partial dot products is not decoupled from the actual dot product computation, leading to performance degradation. This approach is not scalable, as the basic building blocks are specifically designed for King's graph and 4-bit interaction coefficients. The authors believe that this is not a true CIM design and would be more suitable to be regarded as a compute near memory design.  \par The compute in eDRAM presented in Ising-CIM\cite{Shanshan_Ising}\cite{JxCDC} \cite{IGZO_CIM} minimizes data movement by performing XNOR computations within memory. However, the authors assume the interaction coefficients to be unsigned and the design is optimized only for King's Graph. Furthermore, increasing the resolution of interaction coefficients to more than 1 bit is not straight-forward, thus posing scalability challenges. The authors present that 3 cycles are required for a single XNOR compute, thereby degrading the performance. Also, a fully analog CIM approach is sensitive to process variations, reducing accuracy. Similarly, compute in DRAM solutions generally tend to hurt performance and often suffer from power/timing issues \cite{DRAM_PIM}. There is also a more recent work \cite{ABI} that tries to reuse register file in GPUs, by leveraging the fact that register files in GPUs are big enough to accomodate compute in/near-memory 
\par The above compute near/in memory approaches assume that the spins are stationary and involve modifications to the memory array, and this degrades storage density. These approaches pose challenges to performance, scaling and limit the number of CO problems that these approaches can be applied to, for obtaining an optimized solution. However, this work establishes different compute near memory approaches like spin stationary, interaction coefficient stationary and mixed stationary, with specific emphasis on re-configurability, scalability, performance, without modifications to memory array.

\begin{figure}[t]
\centering
\includegraphics[width=\linewidth]{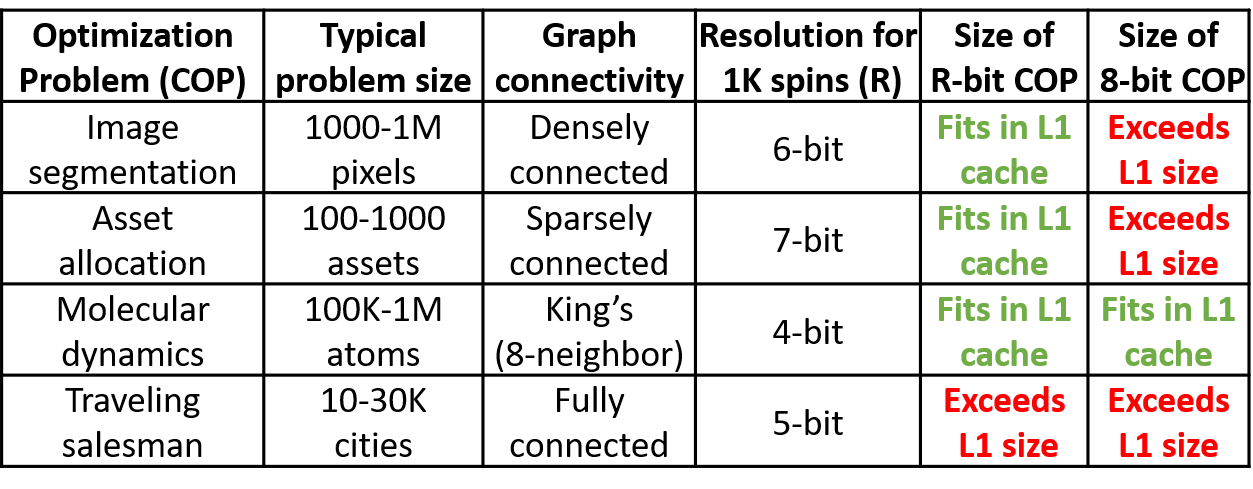}
\vspace{-2em}
\caption{ \textbf{- Real-life COPs have a wide range of problem sizes with varied graph connectivity, varied minimum resolution (4-7 bits) to achieve 90\% accuracy under iso-performance condition. Using a fixed 8-bit IC for all COPs results in additional data movement cost because 8-bit COP overflows when placed in 64KB L1 cache while lower resolution (4-7bit) compute fits inside the L1 cache, motivating a \textbf{\underline{reconfigurable, scalable compute architecture}}.}}
\label{fig:Scalability_problems}
\vspace{-1em}
\end{figure}

\section{ DESIGN GOALS AND MOTIVATION}
\begin{figure}[t]
\centering
\includegraphics[width=\linewidth]{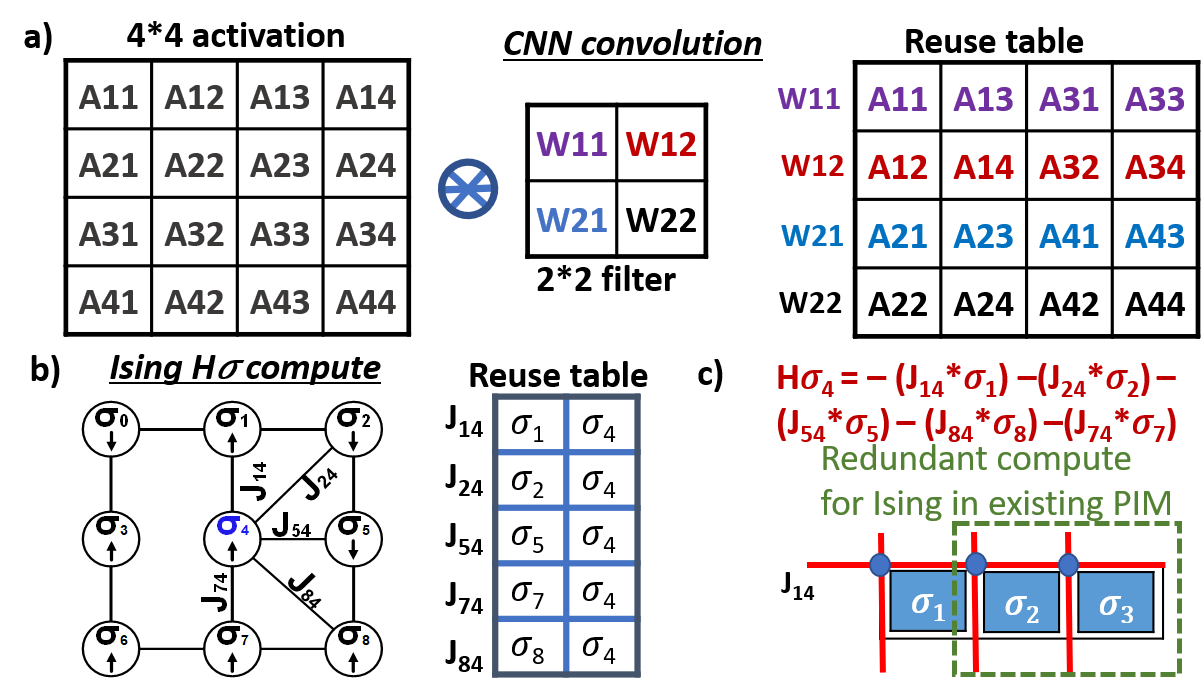}
\vspace{-2em}
\caption{\textbf{a) CNNs offer reuse, as the same weight is shared across multiple activations. b) However, Ising compute for H\textsubscript{$\sigma$} natively offers no reuse as each J\textsubscript{ij} is uniquely mapped between $\sigma$\textsubscript{i} and $\sigma$\textsubscript{j}. c) Memory array mapping for existing PIM design (Ising-CIM) \cite{Shanshan_Ising} shows that although the last 2 columns are computed, they are redundant. This is because J\textsubscript{14} does not interact with $\sigma$\textsubscript{1} and $\sigma$\textsubscript{3} implying that J\textsubscript{14}*$\sigma$\textsubscript{1} and J\textsubscript{14}*$\sigma$\textsubscript{3} are redundant. This redundant compute results in energy degradation. The cause for this degradation is primarily because of the unnecessary discharge of bit cells associated with the redundant compute.}}
\label{fig:ML_Ising_comparison}
\vspace{-1em}
\end{figure}
SACHI is motivated by the weaknesses of prior Ising machines along with an overview of how SACHI overcomes these issues. Earlier architectures need data converters like DACs/ADCs and need specific technologies and/or programming steps coupled to the devices. For instance, ZIV diodes and programmable resistances are required in BRIM, analog charge-sharing with a modified embedded DRAM memory array is required in Ising-CIM. SACHI creates an all-digital architecture with standard components (no specific devices/technology requirements), and can be easily integrated into CMOS SoC. Fig.\ref{Graph_network} summarizes the main features of state-of-the-art Ising machines compared to SACHI. 

\subsubsection{\textbf{ \underline{Repurposability}}}  Domain-specific dedicated accelerators like BRIM/Ising-CIM require frequent CPU-accelerator interaction causing performance/energy overhead. BRIM/Ising-CIM are dedicated because of the difficulty in integrating (i) coupled oscillators made of ZIV diodes and (ii) modified embedded DRAM array into the CPU pipeline. 
\par Instead of a dedicated accelerator, SACHI repurposes the L1 cache when needed, reusing SRAM for in-memory compute with minimal CPU-friendly digital logic.  

\subsubsection{\textbf{ \underline{Scalability}}}  Ising-CIM is explicitly designed for King's graph, without any restriction on the size of COP (detailed in Sec.IV.B). BRIM cannot be scaled for solving large-sized COPs but makes no assumption about the underlying graph. BRIM's scalability is hindered due to the following factors: (i) The number of programmable switches/diodes needed for node interactions scales as O(n\textsuperscript{2}), where n is the number of nodes.  (ii) Ensuring that the capacitance to encode spins does not discharge is crucial to prevent inadvertent spin-flips. Discharge is likely to happen with large problem size. \par SACHI addresses the need to scale to large real-life COPs with diverse connectivity (Fig.\ref{fig:Scalability_problems}), with its unique tuple mapping, tuple-rep property, and storage-array-based updates. 

\subsubsection{\textbf{\underline{Reconfigurability}}}
The compute precision/resolution (R) of Ising-CIM/BRIM is restricted to 2-bit/4-bit. Ising-CIM's restriction arises due to the data mapping that can support only upto 2-bits. BRIM's limitations arise from (i) Challenges in obtaining accurate resistances for higher IC values and the requirement of a configurable DAC for multi-bit R. (ii) Obtaining an 8-bit design involves representing 256 values in 1V range, requiring an infeasibly small voltage resolution for DAC of 1V/256=3mV \par 
SACHI addresses the need to support higher R using mixed-encoding scheme that reduces dot-product into XNOR, enabling high-precision PIM compute without DAC/ADCs. However, a larger R requires more memory space. For 1K spins, achieving 90\% accuracy requires a minimum R of 4-7 bits, depending on the COP as shown in Fig.\ref{fig:Scalability_problems}. SACHI's PIM/near-memory compute makes no assumptions about R and can be reconfigured for any R, without accuracy loss.
\begin{figure}[t]
\centering
\includegraphics[width=\linewidth]{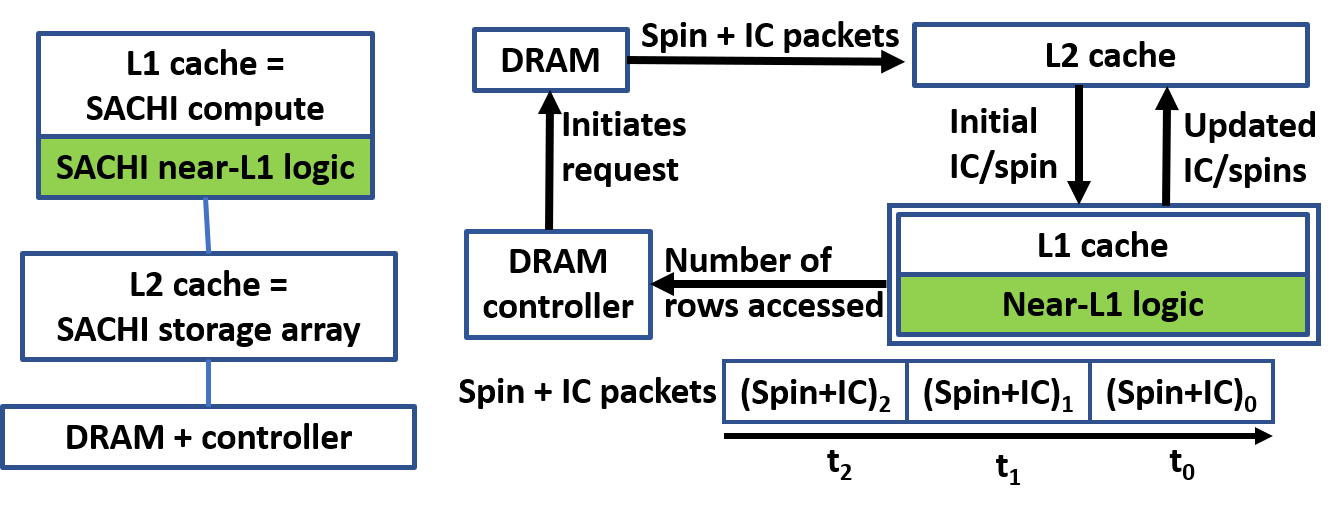}
\vspace{-2em}
\caption{\textbf{\underline{Overview of SACHI}-  SACHI's compute array is mapped onto L1 cache,  storage array onto L2 cache with minimal near-L1 logic (0.3\% of CPU area), enabling easy integration into CPU pipeline. DRAM controller prefetches requests based on the number of uncomputed rows in compute array to minimize data movement cost between DRAM and storage array.}}
\label{DRAM_access}
\vspace{-1em}
\end{figure}

\subsubsection{\textbf{\underline{Reuse}}}
 Reuse for a variable is defined as the number of required computes performed for a variable that is mapped onto (i) a row of compute array in PIM-designs like Ising-CIM, SACHI (ii) a row of coupled oscillator nodes/array in BRIM. BRIM/Ising-CIM offer no reuse. 
\par BRIM has no reuse because every IC mapped onto the coupled oscillator array using ZIV diodes is used in only 1 required compute. Reuse in PIMs is dependent on the algorithm and efficient mapping. 
For example, CNNs inherently exhibit reuse $>$ 1 due to shared weights across activations in a layer, making them suitable for a weight-stationary approach (Fig. \ref{fig:ML_Ising_comparison}a). In contrast, the Ising model for computing H\textsubscript{$\sigma$} (eqn.2) does not naturally offer reuse since each pair of spins and ICs in the graph has a unique mapping.
Fig. \ref{fig:ML_Ising_comparison}b) shows that each J\textsubscript{ij} is uniquely mapped between $\sigma$\textsubscript{i} and $\sigma$\textsubscript{j}, making reuse equal to 1. 
Fig.\ref{fig:ML_Ising_comparison}c) explains the increased energy requirement due to redundant compute. With $\sigma$\textsubscript{1}, $\sigma$\textsubscript{2}, $\sigma$\textsubscript{3} stored in compute array, and J\textsubscript{14} mapped onto a row of compute array, the only required output is J\textsubscript{14}*$\sigma$\textsubscript{1}. However, there are 2 additional redundant computes (J\textsubscript{14}*$\sigma$\textsubscript{2}, J\textsubscript{14}*$\sigma$\textsubscript{3}). The reason for redundant compute is because J\textsubscript{14} does not interact with $\sigma$\textsubscript{2} and $\sigma$\textsubscript{3}. These redundant computes further result in energy overhead due to the unnecessary discharge of bitcells associated with redundant compute. 
Mathematically, an IC tuple (IC[S]~=~$<$IC1, IC2..ICn$>$~;~n$>$1) needs to be formed for each spin, with reuse equal to the number of elements in the formed IC tuple. A design satisfies "IC criterion", if reuse is $>$ 1. Existing PIM design (Ising-CIM) does not satisfy the IC criterion, implying no parallelism across a row of the compute array row. 
\par SACHI \cite{SACHI} uses a reuse-aware data-stationary compute strategy to improve reuse across all elements in a row of compute array, improving parallelism and energy efficiency.  

\begin{figure}[t]
\centering
\includegraphics[width=\linewidth]{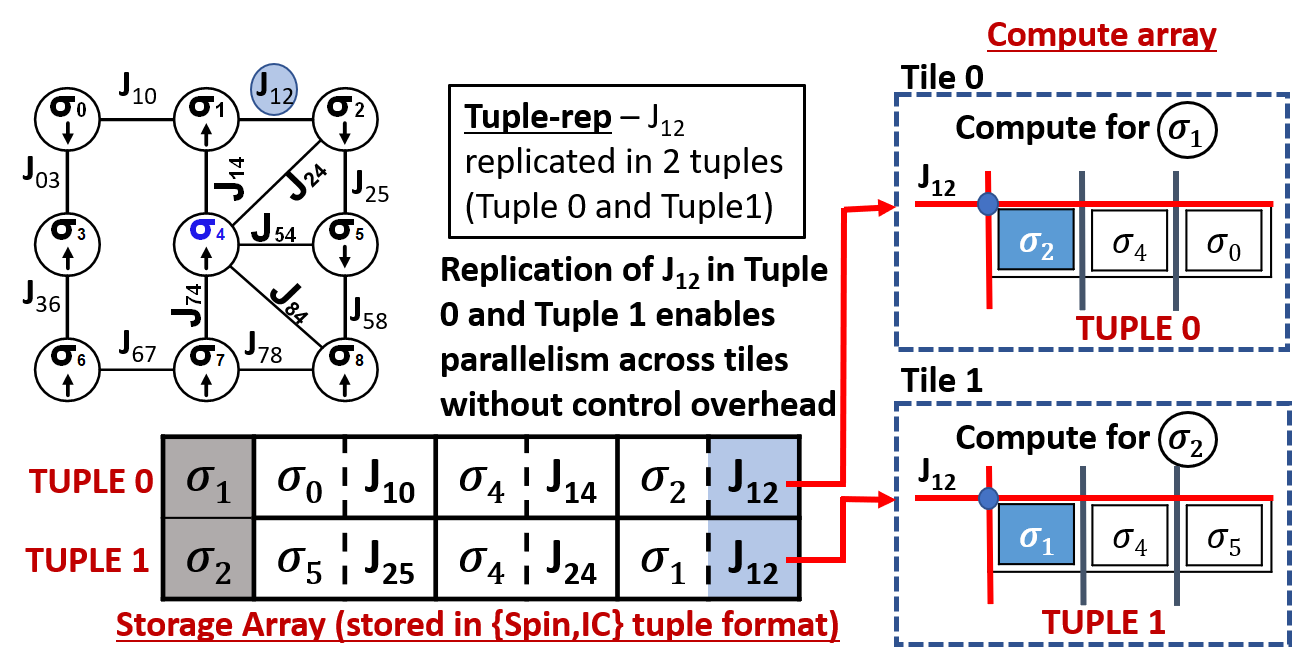}
\vspace{-2em}
\caption{ \textbf{\underline{Illustration of tuple mapping strategy} - a) Storage array organized as tuples, with each consisting of the connected spins and ICs for a given spin. Ising compute operates at the abstraction level of these tuples, thereby not requiring visibility of the incoming graph helping achieve scalability to any graph type. b) Tuple-rep ensures 1:1 mapping between a storage array tuple and a compute tile row, enabling the partitioning of the graph into sub-graphs, with each sub-graph computed independently in different tiles without any control overhead}}
\label{Scalability_tuple}
\vspace{-1em}
\end{figure}

\section{ THE SACHI ARCHITECTURE}
We describe the SACHI architecture, which combines elements of re-purposability, scalability, reconfigurability, and reuse-aware data-stationary near-memory compute, to realize a high-performance, energy-efficient Ising machine.

\subsection{Mathematical intuition of proposed architecture}

Mapping onto memory array should be optimized in such a way that it makes different arithmetic computations feasible. Machine learning accelerators utilize the weight stationary approach for computing dot product in memory. The major reason behind using this approach, as opposed to the activation stationary approach, is that this ensures maximum reuse of weights(W), as all the activations(A) in a layer share the same set of weights. Mathematically, the reuse can be defined as the feasibility of forming an activation tuple(A) of more than one activation, wherein each activation tuple is indexed by the weight shared across all activations in the tuple. The number of elements in the activation tuple is a measure of reuse, measured as follows:
\begin{equation}
    A[W]=<A1,A2..An> where n>1
\end{equation}
In the case of Ising accelerators, the mapping should be in such a way that it makes computation of \ref{eq:h_sigma} feasible. Intuitively, \ref{eq:h_sigma} for a target spin, can be explained as the accumulation of the dot product of neighboring spin and the interaction coefficient between the target and neighboring spins. There is a unique mapping between a pair of spin(S)s and the interaction coefficient(IC) between the spins, in case of computing equation \ref{eq:h_sigma}.  Re-using of spins/interaction coefficients requires in-depth analysis of the parameters that needs to be stationary in the compute array.
Thus, the mapping equation can be written as follows:
\begin{equation}\label{eq:IC_criterion}
     IC[S]=<IC1,IC2..ICn> where n>1
\end{equation}
 which implies that multiple ICs share the same S in computing \ref{eq:h_sigma}. This equation will be regarded as IC criterion in this paper.
 An equally valid representation would be to make sure that multiple spins share the same IC(S criterion in this paper) 
 \begin{equation}\label{eq:s_criterion}
     S[IC]= <S1,S2..Sn> where n>1
 \end{equation}
 \begin{figure}[t]
\centering
\includegraphics[width=\linewidth]{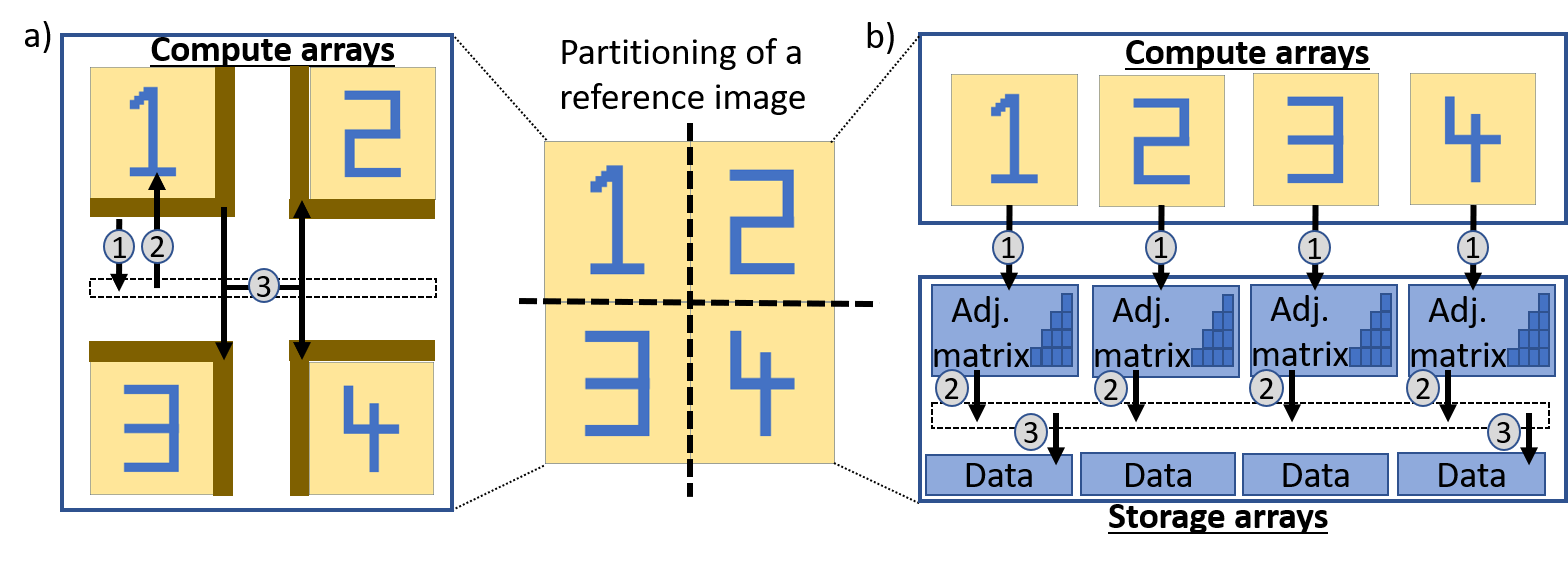}
\vspace{-2em}
\caption{ \textbf{\underline{Scalability to any graph size} - a) Ising-CIM approach: duplicate edge cells (dark brown) onto adjacent CIM arrays. 1 indicates updated spin value computed in the array. 2 indicates writeback of the same updated spin value for non-edge cells. 3 indicates the broadcast of updated spin value in the case of edge cells (duplication) to adjacent CIM arrays. This approach minimizes interaction between compute arrays to only edge cells, but uses King's Graph properties to accomplish this and is performance-inefficient b) SACHI approach:  store the adjacency matrix and  buffer the updated spins of relevant tuples by reading the adjacency matrix. 1 indicates updated spin value computed in the array. 2 indicates read of the adjacency matrix. 3 indicates update of relevant tuples in storage array. This is graph-type agnostic and performance-efficient}}
\label{Scalability_size}
\vspace{-1em}
\end{figure}
It is to be noted that spins are always binary. Identifying a solution to IC criterion can be simplified as finding a mapping strategy that would enable storage of interaction coefficients onto memory array and mapping spins onto Word line/Bit line, as these are shared by bitcells in a row/column. On the other hand, finding a solution to S criterion can be described as mapping an IC onto the Word line/Bit line, so that multiple spins stored in the memory array share the same IC. However, because interaction coefficients are of multiple bits, mapping onto word line/bit line would require a bit-serial mapping, implying that the different bits are mapped during different cycles with minimal parallelism. Thus, a mapping strategy that satisfies IC criterion would complete all computations in one iteration, while an efficient S criterion (without adding analog circuits) satisfying mapping would take several iterations to compute. Thus, for achieving better performance, IC criterion needs to be satisfied for processing.  

\subsection{Repurposable architecture} SACHI repurposes modern-day CPU SoC components to accelerate Ising Hamiltonian computation (H\textsubscript{$\sigma$}) with minimal area overhead. The components used include DRAM, L2 cache, CPU, and a repurposed L1 cache, utilizing 8T SRAM bitcells (Fig.\ref{DRAM_access}). The storage array is mapped onto L2 cache, while the compute array is mapped onto L1 cache, with additional near L1 peripheral logic occupying only 0.3\% of AMD's Zen3 CPU area. The data flow in SACHI is as follows: For spins and ICs that fit on-chip, DRAM is accessed once at the beginning to fill the storage arrays, and no further re-access is needed. The storage array (L2 cache) acts as a buffer for storing the initial Ising graph transferred from DRAM, while the compute array (L1 cache) performs H\textsubscript{$\sigma$} computations using the spins/ICs from the storage array and writes the updated spins back onto the storage array. In cases where variables and ICs do not fit on-chip, multiple rounds of computations are required to optimize large graphs, necessitating rewriting of the compute and storage arrays. To optimize this long latency operation, we use a prefetching approach that anticipates potential accesses. CIM accesses have structured and predictable address patterns, unlike regular memory access. We access the rows in the compute array top-to-bottom in successive cycles, and a counter in the DRAM controller tracks the number of remaining rows to be accessed. When the count reaches a threshold (meeting DRAM-to-storage + storage-to-compute array data movement latency), a prefetch request is initiated to ensure the timely arrival of DRAM-requested data.


\subsection{Scalability using tuple mapping} \subsubsection{\textbf{\underline{Scalability to all graphs}}} SACHI operates on the incoming graph at the abstraction level of tuples, enabling H\textsubscript{$\sigma$} compute to be incognizant of the graph connectivity. This unique feature enables  SACHI to perform Ising computations for all graphs, irrespective of their connectivity. Fig.\ref{Scalability_tuple}a) illustrates our tuple mapping strategy. Each row in the storage array is a tuple for a particular spin, consisting of the neighboring spin states, the connecting ICs, and the external magnetic field (Fig.\ref{Scalability_tuple}a) 
Furthermore, the same IC/spin is present in more than one row, called the \textbf{"tuple-rep"} property (Fig.\ref{Scalability_tuple}b). For instance, J\textsubscript{12} is an entry present in the tuples of both $\sigma$\textsubscript{1} and $\sigma$\textsubscript{2}. This enables the compute for H\textsubscript{$\sigma1$} and H\textsubscript{$\sigma2$} to be independent of each other by ensuring 1:1 mapping between a tuple in the storage array, and a row in the compute array, enabling partitioning of large graphs into subgraphs. If not for tuple-rep (J\textsubscript{12} was present only in $\sigma$\textsubscript{1}'s tuple), H\textsubscript{$\sigma$} compute for $\sigma$\textsubscript{2} introduces an interdependency of rereading the storage array for obtaining J\textsubscript{12} from the tuple corresponding to $\sigma$\textsubscript{1}, causing performance bottlenecks with control overhead.

\subsubsection{\textbf{\underline{Scalability to any graph size}}} To efficiently solve large COPs, reducing inter-CPU core interactions is crucial. For PIM designs, this involves minimizing interactions between sub-arrays of compute array, while being graph-type agnostic, and extending the same philosophy to reduce inter-core interactions. Firstly, unlike deep-neural networks, where the output of one layer feeds as input to another, requiring data movement, Ising model does not incur any "layerwise" data movement. 
The only required minimal data movement is for spin-updates. Therefore, the algorithm inherently requires fewer inter-array interactions. In \textbf{Ising-CIM}, when graphs are partitioned, spins on the edge of the partition are duplicated across 2 adjacent CIM arrays. Non-edge cells perform local spin updates in eDRAM (2 in Fig.\ref{Scalability_size}a) based on the computed updated spin value (1 in Fig.\ref{Scalability_size}a). Edge cells undergo a local read-modify-write(update) based on the broadcasted updated spin value from adjacent CIM arrays (3 in Fig.\ref{Scalability_size}a). Although only edge cells necessitate interaction between adjacent arrays, this approach has several drawbacks. Execution time is dependent on a)cycles per iteration (CPI) and b)number of iterations (IT). With respect to CPI,  the local update in Ising-CIM hinders Ising compute performance, making each compute a 2-cycle operation (1 each for compute and update), because of read-write conflict. SACHI overcomes this, making compute 1-cycle operation, as the data movement from compute-storage array is overlapped with useful compute in compute array. Furthermore, there is no read-write conflict, as the write happens to a separate array. Therefore, Ising-CIM has 2x CPI compared to SACHI. With respect to IT, local update leads to performance gain only in large-sized COPs. For instance, in localized King's Graph, there is a minor performance gain of 0.1x, as opposed to 1.8x in a complete graph for 1M spin configuration. The gain is with respect to performing storage array based update, similar to SACHI. Therefore, in Ising-CIM, it is beneficial to retain the original values and reap benefits from improved CPI. In SACHI, this happens naturally for COPs that require re-write to compute array. A portion of the storage array stores the adjacency matrix, and is read (2 in Fig.\ref{Scalability_size}b) to identify the relevant tuples containing the incoming spin, allowing updates to the storage array (3 in Fig.\ref{Scalability_size}b) only for the relevant tuples. This ensures that when the compute array is re-written, few tuples already have the updated spin values Therefore, SACHI provides the ideal middle-ground mimic-ing local update behavior for fast convergence, when necessary, while providing improved performance due to 1-cycle compute+update operation.
\begin{figure}[t]
\centering
\includegraphics[width=\linewidth]{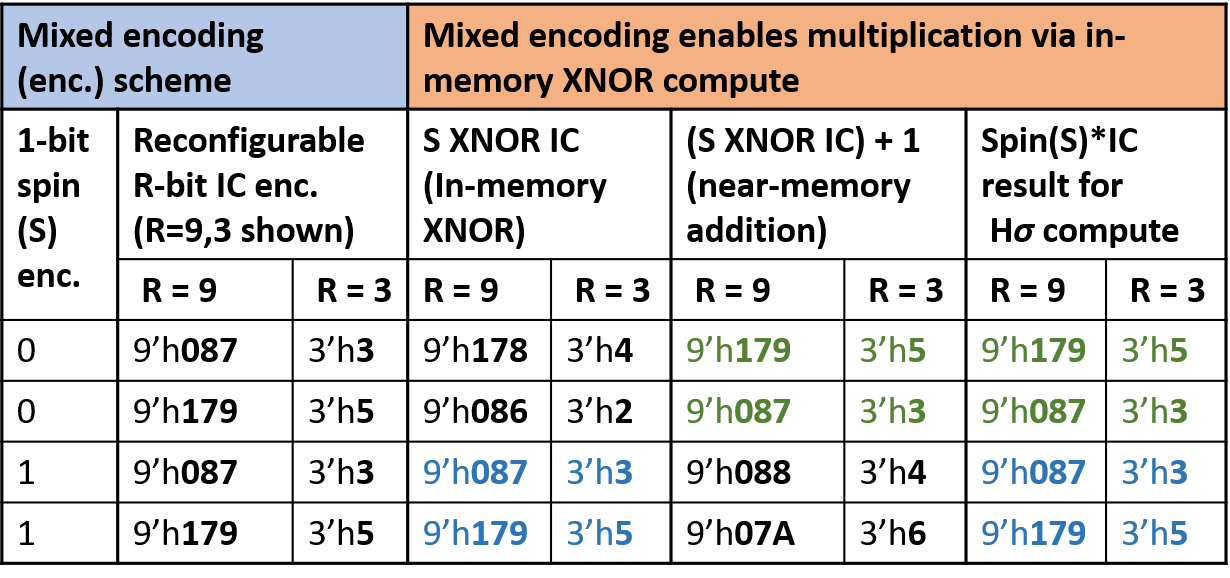}
\vspace{-2em}
\caption{ \textbf{\underline{SACHI's reconfigurability} achieved using \textbf{mixed encoding scheme, with -1/+1 spins stored as 0/1, ICs encoded in 2's complement form to enable in-memory XNOR for dot product between J\textsubscript{ij} and $\sigma$\textsubscript{j} without DAC/ADC (unlike BRIM). (S XNOR IC)/((S XNOR IC)+1) is computed to enable multi-bit signed IC dot product (unlike Ising-CIM). 9-bit J\textsubscript{ij}= 135 (9'h087), -135 (9'h179) and 3-bit J\textsubscript{ij}= 3 (3'h3), -3 (3'h5) product with $\sigma$\textsubscript{j}=1 (0),-1 (1) is shown}
 }}
\label{Reconfigurability_encoding}
\vspace{-1em}
\end{figure}
\begin{figure*}[t!]
\centering
\includegraphics[width=\linewidth]{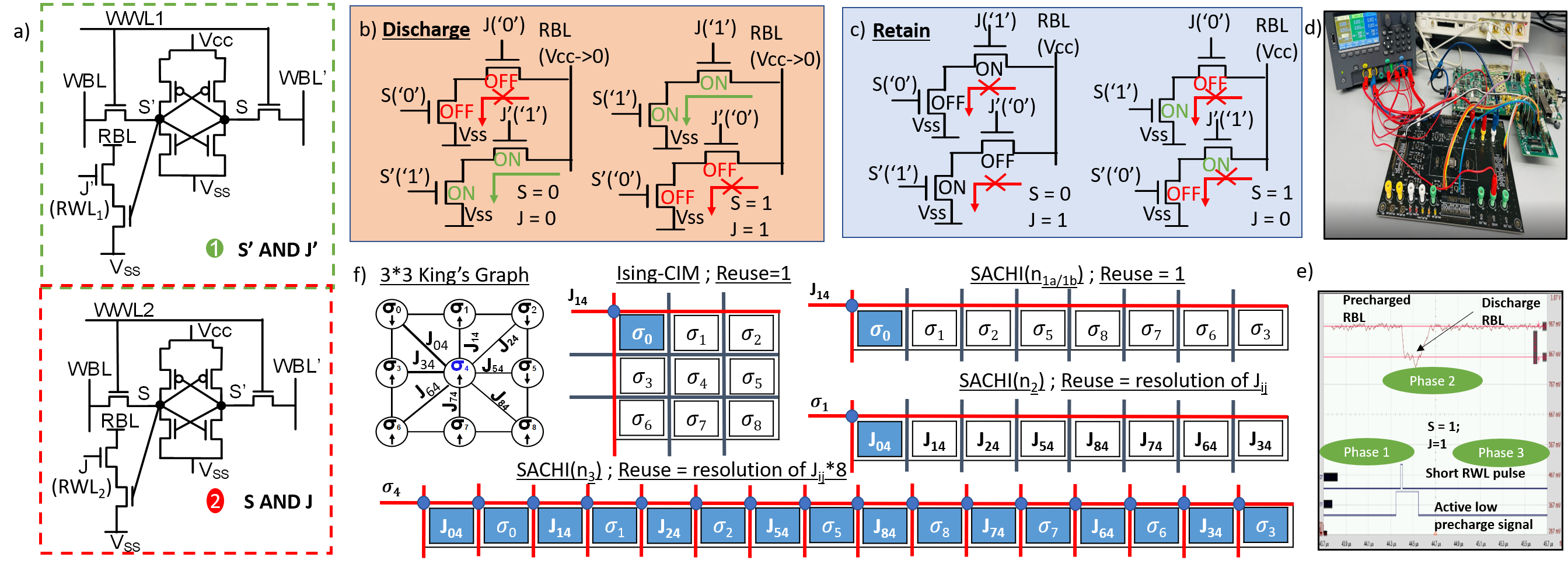}
\vspace{-2em}
\caption{\textbf{{SACHI's use of in-memory XNOR} - a) 2 8T SRAM bitcells in the same column storing complimentary values of spin/IC with the bitcells computing S'\&J', S\&J b) RBL discharge c) RBL retain for XNOR compute with \textbf{no memory array modifications/requirement of DAC and ADC} d) In-memory XNOR test-structure in a silicon prototype e) Oscilloscope capture confirming bitline discharge for S=1 XNOR J=1 f) \textbf{\underline{SACHI reuse in different stationarities illustrated with King's graph} -  SACHI(n\textsubscript{1a}) has reuse=1, as in Ising-CIM as J\textsubscript{14} is used only for multiplying with $\sigma$\textsubscript{0}. SACHI(n\textsubscript{2}),  SACHI(n\textsubscript{3}) employ IC,  mixed stationary design respectively to improve reuse to J\textsubscript{ij}  resolution, (J\textsubscript{ij} resolution*8) respectively}  }}
\label{8T_SRAM_compute}
\vspace{-1em}
\end{figure*}

\par \subsubsection{\textbf{\underline{Scalability to any graph size}}} To efficiently solve large COPs, reducing inter-CPU core interactions is crucial. For PIM designs, this involves minimizing interactions between sub-arrays of compute array, while being graph-type agnostic, and extending the same philosophy to reduce inter-core interactions. Firstly, unlike deep-neural networks, where the output of one layer feeds as input to another, requiring data movement, Ising model does not incur any "layerwise" data movement. 
The only required minimal data movement is for spin-updates. Therefore, the algorithm inherently requires fewer inter-array interactions. In \textbf{Ising-CIM}, when graphs are partitioned, spins on the edge of the partition are duplicated across 2 adjacent CIM arrays. Non-edge cells perform local spin updates using the local write drivers in eDRAM (2 in Fig.\ref{Scalability_size}a) based on the computed updated spin value (1 in Fig.\ref{Scalability_size}a). Edge cells undergo a local read-modify-write based on the broadcasted updated spin value from adjacent CIM arrays (3 in Fig.\ref{Scalability_size}a). Although only edge cells necessitate interaction between adjacent arrays, this approach has several drawbacks. Firstly, the local update for non-edge cells causes the loss of the original spin value before completing an iteration. While this approach is scalable to large-sized King's Graph due to King's Graph's localized interaction nature, wherein the original spin value is not used again in the same iteration, it cannot be extended to other complex graphs with non-local interaction where the original spin value is reused sometime later. Secondly, the read-modify-write operation hinders Ising compute performance, making each compute a 2-cycle operation (1 each for compute and update). \textbf{SACHI} enables scalability by repurposing the storage array for writing the computed updated spin values (1 in Fig.\ref{Scalability_size}b), ensuring that the original spin value remains intact in the compute array, without requiring interaction between compute arrays. A portion of the storage array stores the adjacency matrix for connectivity information. While H\textsubscript{$\sigma$} compute is incognizant of graph connectivity, the update needs to be aware of the connectivity. This matrix is read (2 in Fig.\ref{Scalability_size}b) to identify the relevant tuples containing the incoming spin, allowing updates to the storage array (3 in Fig.\ref{Scalability_size}b) only for the relevant tuples. Furthermore, data movement latency between storage and compute arrays is hidden by performing useful compute in the compute arrays.

\subsection{Reconfigurability using mixed encoding scheme} 
\subsubsection{\textbf{\underline{Mixed-encoding}}} SACHI uses a mixed-encoding scheme, wherein +1/-1 spins are encoded as 1/0 and ICs are represented using 2's complement form to enable PIM dot-product for signed multi-bit ICs without DAC/ADC. The dot product (Fig.\ref{Reconfigurability_encoding}) between J\textsubscript{ij} and $\sigma$\textsubscript{j} for eqn.2 is simplified as an XNOR operation, leveraging the binary nature of spins, enabling PIM without any array modifications: 

\begin{equation}\label{dot_product}
    J\textsubscript{ij}*\sigma\textsubscript{j}=
    \begin{cases}
        J\textsubscript{ij}  $XNOR$ \sigma\textsubscript{j}, & \sigma\textsubscript{j} > 0 \\
        J\textsubscript{ij} $XNOR$ \sigma\textsubscript{j} + 1, & \sigma\textsubscript{j} < 0 \\
    \end{cases}
\end{equation}
\begin{figure*}[t!]
\centering
\includegraphics[width=\linewidth]{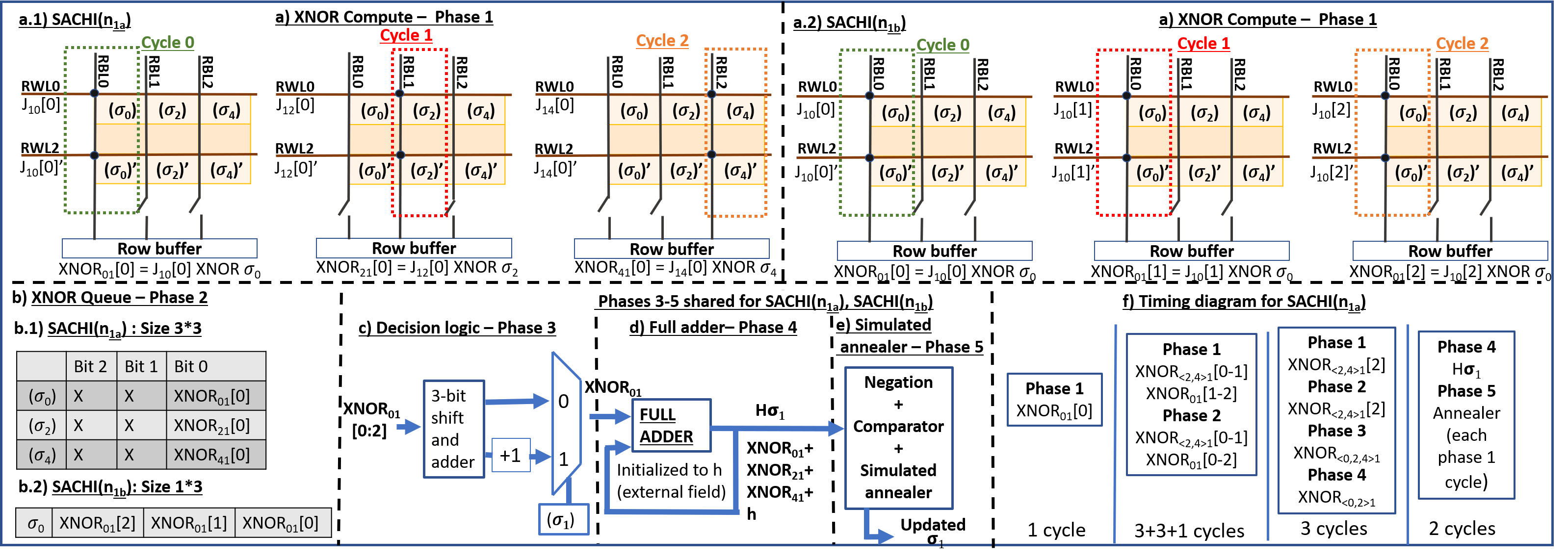}
\vspace{-2em}
\caption{\textbf{ a.1) SACHI(n\textsubscript{1a}) a.2) SACHI(n\textsubscript{1b}) for graph in Fig.\ref{Graph_type} with 3-bit J\textsubscript{ij}. SACHI(n\textsubscript{1a}) computes i\textsuperscript{th}bit of all ICs before proceeding to the (i+1)\textsuperscript{th} bit (reuse=1). SACHI(n\textsubscript{1b}) computes all bits of an IC before proceeding to the next IC (reuse=1). Only 1 column is highlighted to indicate reuse of 1. \textbf{b) Bitwise XNOR} requires XNOR queue to queue the different bits computed \textbf{b.1)} SACHI(n\textsubscript{1a}) requires more number of entries in the queue than \textbf{b.2)} SACHI(n\textsubscript{1b}) because of Phase 3 preferring SACHI(n\textsubscript{1b}) order of compute \textbf{c) Phase 3} performs shift and add among all bits of XNOR compute \textbf{d) Phase 4} adds the partial XNOR products and \textbf{e) Phase 5} for simulated annealing. Phases 3-5 are the same for SACHI(n\textsubscript{1a}) and SACHI(n\textsubscript{1b}) \textbf{f) Timing diagram} for SACHI(n\textsubscript{1a}) showing the overlap between phases 1-4 for H\textsubscript{$\sigma$} compute
}}
\label{fig:ISCA(n1a_b)}
\vspace{-1em}
\end{figure*}

\subsubsection{\textbf{\underline{In/near-L1 compute}}} The XNOR operation is performed by activating multiple rows simultaneously of the compute array. To accomplish this, L1 cache, typically made of an 8T SRAM bitcell (unchanged from what is used in modern-day CPUs for L1 caches \cite{8T_L1}) with decoupled read and write ports (RWL, RBL, WWL, WBL - read/write word/bit lines) is employed.  
\cite{8T_SRAM} (Fig. \ref{8T_SRAM_compute}a). This bitcell operates in two modes: (i) regular read/write mode and (ii) Ising compute mode. In the regular mode, data is written/read using WBL/RBL by enabling WWL/RWL, respectively. RWL is repurposed for computation during Ising compute mode. Logical AND between input (J) and stored value (S) is achieved by driving RWL\textsubscript{2} with J. For XNOR (S AND J) OR (S' AND J') operation, S' is stored in a different bitcell in the same column, and RWL\textsubscript{1} is driven with J' (Fig.\ref{8T_SRAM_compute}a). RBL discharges when either (S AND J) or (S' AND J') is high, indicating an XNOR value of 1 (Fig. \ref{8T_SRAM_compute}b). RBL retains its precharged value when both (S AND J) and (S' AND J') are low, indicating an XNOR value of 0 (Fig. \ref{8T_SRAM_compute}c). The oscilloscope capture for a prototype in TSMC 65nm technology process shows RBL discharge for a single column of SRAM array (size 100*100) (Fig. \ref{8T_SRAM_compute}d,e).
The waveform for a selected bitcell storing '1' is summarized as: Phases 1/3 - Precharge: An active low precharge signal charges the RBL to 1V.
Phase 2 - Compute: A short RWL pulse indicates the incoming value ('1'). Since both the storage node and the incoming RWL are '1', RBL discharges, resulting in XNOR value of 0 (Fig. \ref{8T_SRAM_compute}b illustrates the discharge path). Dot product accumulation is performed using full adders situated near memory. This enables (i) parallel execution of PIM XNOR and accumulation, and (ii) enhanced accuracy due to the reduced susceptibility to process variations in the digital full adder.  Accurate in/near-L1 design without assumptions about R makes SACHI support any R-bit compute. 
\subsection{Sensitivity to memory/temperature} 
All these compute operations are shown using SRAM \cite{GCN}\cite{SPARK}\cite{NEM_GNN_arxiv} as the underlying bitcell at room temperature, it is important to note that there is nothing fundamental that prevents usage of other emerging non-volatile memory technology bitcells like Ferroelectric Field Effect Transistors (FeFET) \cite{fefet}, Resistive Random access memory \cite{RRAM_1}\cite{RRAM_2}\cite{NVM_Raman}\cite{RRAM_cache}\cite{UT_Thesis}\cite{SPARK_arxiv}, etc. Similarly, this operation can be performed at cryogenic temperature using flop-array based designs \cite{UTBB_SOI} or emerging technologies like Josephson Junction Field Effect Transistors \cite{JJFET}\cite{Cryo_arxiv}.  

\par 
\subsection{Reuse-aware data stationary compute}
We propose near-memory architectures that progressively achieve higher reuse: (i) spin stationary (SACHI (n\textsubscript{1a})), (SACHI(n\textsubscript{1b})), (ii) IC stationary (SACHI(n\textsubscript{2})), (iii) Mixed stationary (SACHI(n\textsubscript{3})). (Fig.\ref{8T_SRAM_compute}f) illustrates an overview of the three methods with King's graph. 

\begin{algorithm}[t]
\caption{Row decoding for spin stationary design}
\begin{algorithmic}[1]

\Procedure{DECODE}{$NumSpins, n, N$}\label{alg:b}

\If{$\text{COMPUTE AND SACHI}(n_{1a})$}

    \For{$s=0; s<NumSpins; s++$}
        \For{$J=0; J<n; J++$}
            \For{$m=0; m<N; m++$}
                \State $RWL[s] \gets J_{ij}[s][J][m]$
                \State $RWL[s+NumSpins] \gets \neg J_{ij}[s][J][m]$
                \State $SAEn[m] \gets 1$
            \EndFor
        \EndFor
    \EndFor

\ElsIf{$\text{COMPUTE AND SACHI}(n_{1b})$}

    \For{$s=0; s<NumSpins; s++$}
        \For{$m=0; m<N; m++$}
            \For{$J=0; J<n; J++$}
                \State $RWL[s] \gets J_{ij}[s][J][m]$
                \State $RWL[s+NumSpins] \gets \neg J_{ij}[s][J][m]$
                \State $SAEn[m] \gets 1$
            \EndFor
        \EndFor
    \EndFor

\EndIf

\EndProcedure

\end{algorithmic}
\end{algorithm}

\subsubsection{\textbf{\underline{SACHI(n\textsubscript{1a}) = Spin   stationary design}}} 

This design involves storing spins ($\sigma$) onto the compute array and mapping ICs onto RWL in a bit-serial manner. This approach, illustrated in Fig.~\ref{fig:ISCA(n1a_b)} allows sharing a J\textsubscript{ij} bit across a row of bitcells in the compute array, which is organized as tiles and filled in order (successive spins in the same tile) without interleaving. 
\par The compute is as follows:
(i) In \textbf{phase 1}, the sharing of J\textsubscript{ij} across a row of spins leads to redundant XNOR computes. The read-out of redundant dot products is blocked by disabling the bit select of each column~(Fig.~\ref{fig:ISCA(n1a_b)}a.1), giving a throughput of 1 XNOR compute every cycle. As you will see, the hardware requirements are influenced by the order of XNOR compute. In SACHI(n\textsubscript{1a}), XNOR of r\textsuperscript{th} bit of all the neighboring spins is performed before computing XNOR of (r+1)\textsuperscript{th} bit (Fig.~\ref{fig:ISCA(n1a_b)}a.1). 
This compute order necessitates a storage buffer (called XNOR queue) to store XNOR of individual bits in \textbf{phase 2}(Fig.~\ref{fig:ISCA(n1a_b)}b.1). The minimum size of the queue equals (number of neighboring spins * (resolution of J\textsubscript{ij}+1). In \textbf{phase 3} (Fig.~\ref{fig:ISCA(n1a_b)}c), the computed XNOR values generate partial products by shifting and adding individual bits. Additionally, a decision is made based on $\sigma$\textsubscript{j} to choose between XNOR and (XNOR + 1) after XNOR'ing all the bits of J\textsubscript{ij} with $\sigma$\textsubscript{j}. 
In \textbf{phase 4}, the full adder, which is initialized to the external magnetic field,  accumulates the partial dot products for each spin (Fig.~\ref{fig:ISCA(n1a_b)}d). In \textbf{phase 5} (Fig.~\ref{fig:ISCA(n1a_b)}e), the computed sum is negated to obtain H\textsubscript{$\sigma$} and is passed through annealer.

 \par 
The time to compute H\textsubscript{$\sigma$} for an R-bit IC with N neighbors for a specific spin is in O(R*N). 
Notably, N cycles are required to fill one column in the XNOR queue, repeated (R-1) times to fill (R-1) columns, and 1 cycle is needed to obtain the first R-bit value with shift and add. This results in ((R-1)*N+1) cycles before phase 3 becomes active (Fig.\ref{fig:ISCA(n1a_b)}f). During this time, phases 3-5 are idle, referred to as "idle time". The reuse is 1,
as every J\textsubscript{ij} bit fetched from the storage array is used in 1
XNOR compute, not satisfying the IC criterion. 

\subsubsection{\textbf{\underline{SACHI(n\textsubscript{1b}) = Optimized SACHI(n\textsubscript{1a})}}} SACHI(n\textsubscript{1a}) can be improved in terms of (i) effective resource utilization by minimizing the "idle time" (ii) reducing the overhead of XNOR queue and (iii) better interleaving of tiles. 
\par 
To tackle the first two issues, we modify the order of XNOR computes to achieve a directed throughput, allowing earlier activation of phases 3-5. Additionally, we store adjacent rows of the storage array in adjacent tiles to improve interleaving. 

\par The compute involves mapping RWL onto successive bits of a specific J\textsubscript{ij} in consecutive cycles to perform XNOR during \textbf{phase 1}. This differs from SACHI(n\textsubscript{1a}), where this mapping is delayed by N cycles. In other words, all bits of a particular J\textsubscript{ij} are XNOR-computed before moving on to the next J\textsubscript{ij}, e.g.  XNOR\textsubscript{01}$[$0-2$]$ is followed by XNOR\textsubscript{21}$[$0-2$]$. 
This order is evident as the bitline select of column 0 is turned ON three times (Fig.~\ref{fig:ISCA(n1a_b)}a.2), reducing the XNOR queue size to 1 row with R columns (Fig.~\ref{fig:ISCA(n1a_b)}b.2) in \textbf{phase 2}. \textbf{Phases 3-5} are same as SACHI(n\textsubscript{1a}). Additionally, phase 3 is initiated earlier than SACHI(n\textsubscript{1a}) since J\textsubscript{ij} bits are computed earlier to shift and add.

\par The compute time for H\textsubscript{$\sigma$} is in O(R*N). However, the idle time is reduced from (R-1)*(N) to R cycles 
SACHI(n\textsubscript{1b}) does not satisfy the IC criterion, as reuse is 1 (as every J\textsubscript{ij} bit fetched is used only in 1 compute). 


 \begin{algorithm}[b]
\caption{Row decoding for IC stationary design}
\begin{algorithmic}[1]

\Procedure{DECODE}{$NumSpins, n, N$}\label{alg:c}

\If{$\text{COMPUTE}$}

    \For{$s=0; s<NumSpins; s++$}

        \For{$J=0; J<(n \times N); J+=n$}

            \State $SAEn[J] \text{ to } SAEn[J+n-1] \gets 1$

            \For{$k=0; k<n; k++$}

                \State $RWL[s] \gets \sigma_J[s][J+k]$

                \State $RWL[s+NumSpins] \gets
                \neg \sigma_J[s][J+k]$

            \EndFor

        \EndFor

    \EndFor

\EndIf

\EndProcedure

\end{algorithmic}
\end{algorithm}

 \begin{figure}[t]
\centering
\includegraphics[width=\linewidth]{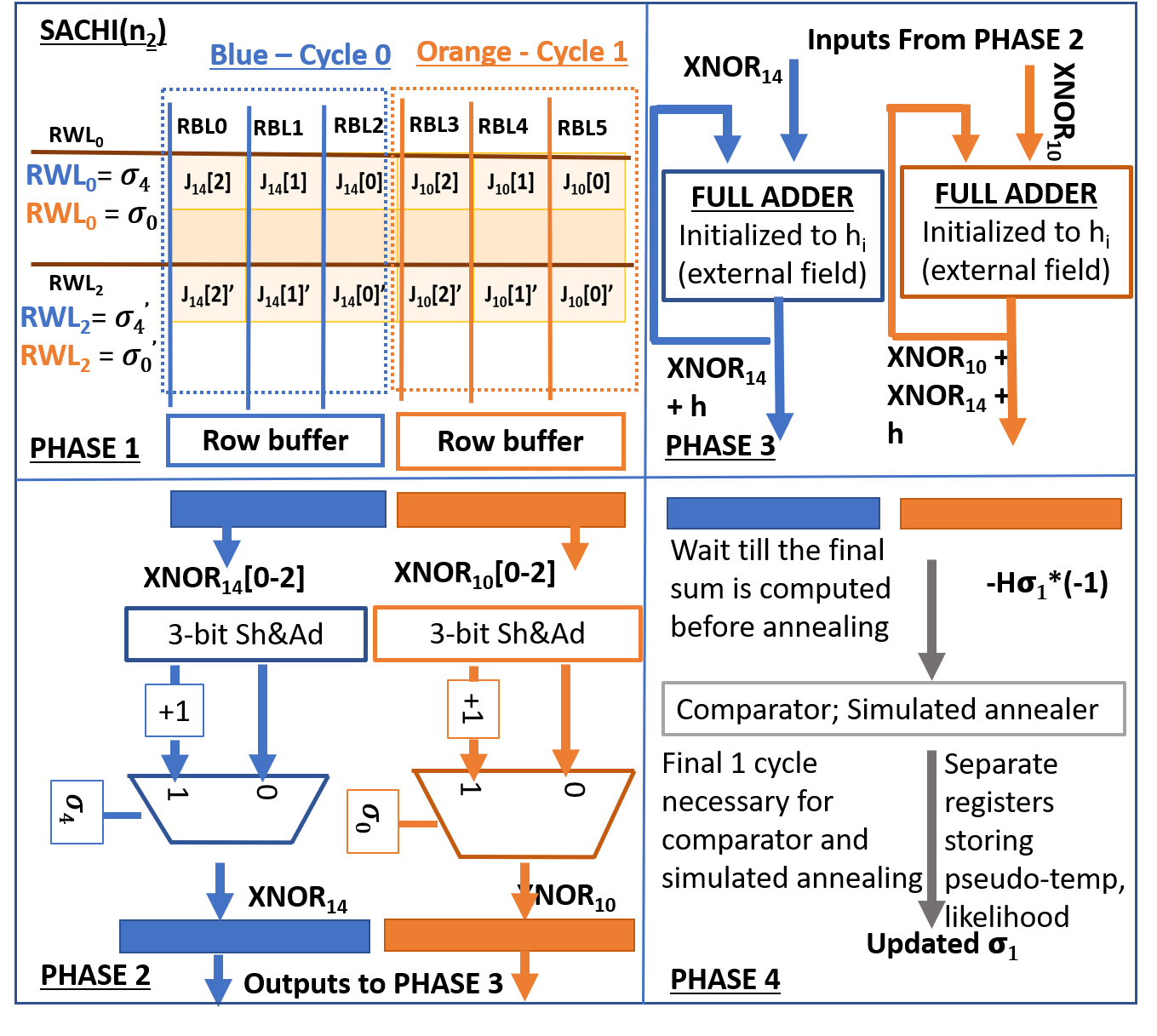}
\vspace{-2em}
\caption{\textbf{\underline{SACHI(n\textsubscript{2}): IC stationary design} performing XNOR across all J\textsubscript{ij} bits in a single cycle in Phase 1, improving maximum reuse to the resolution of J\textsubscript{ij} (3 columns highlighted for 3-bit J\textsubscript{ij}), eliminating XNOR queue in Phase 2. The decision logic, full adder, and simulated annealing logic are shifted by a cycle compared to SACHI(n\textsubscript{1})}}
\label{fig:ISCA(n2)}
\vspace{-1em}
\end{figure}\subsubsection{\textbf{\underline{SACHI(n\textsubscript{2}) = Interaction Coefficient 
stationary design}}} 
SACHI(n\textsubscript{1a/b}) suffer from (i) low SRAM throughput (ii) low reuse/performance (iii) XNOR queue overhead. 
 To address these issues, SACHI(n\textsubscript{2}) computes multiple XNOR operations in parallel by storing ICs (J\textsubscript{ij}) onto compute array and mapping spins onto RWL. For every i\textsuperscript{th} row's RWL mapped to a spin, [i+num spins]\textsuperscript{th} row's RWL is mapped to the inverted spin value for XNOR compute. 
 \par The compute illustrated in Fig.~\ref{fig:ISCA(n2)} proceeds as follows: In \textbf{phase 1}, multiple bitline selects are turned ON to read the computed data from multiple columns so that all bits of a particular XNOR partial product between J\textsubscript{ij} and S\textsubscript{j} are obtained in 1 cycle. Thus, the throughput of the SRAM array is improved from 1 in SACHI(n\textsubscript{1a/b}) to R in SACHI(n\textsubscript{2}), thereby eliminating the XNOR queue. In \textbf{phase 2}, the decision between XNOR and XNOR+1 is taken based on the target spin.  In \textbf{phase 3}, full adder accumulates the partial dot products. 
\par H\textsubscript{$\sigma$} compute time is reduced from O(N*R) in SACHI(n\textsubscript{1}) to O(N) in SACHI(n\textsubscript{2}) and the reuse is equal to resolution (R) of SACHI(n\textsubscript{1a/b}), as a single spin is shared by multiple bit-cells, thus satisfying the IC criterion.
 



\begin{algorithm}[t]
\caption{Row decoding for mixed stationary design}
\setlength{\textfloatsep}{0pt}

\begin{algorithmic}[1]

\Procedure{DECODE}{$NumSpins$}\label{alg:d}

\If{$\text{COMPUTE}$}

    \For{$s=0; s<NumSpins; s++$}

        \State $RWL[s] \gets \sigma_i[s]$

        \State $RWL[s+NumSpins] \gets \neg \sigma_i[s]$

        \State $SAEn \gets 1$

    \EndFor

\EndIf

\EndProcedure

\end{algorithmic}
\end{algorithm}

\begin{figure}[t]
\centering
\includegraphics[width=\linewidth]{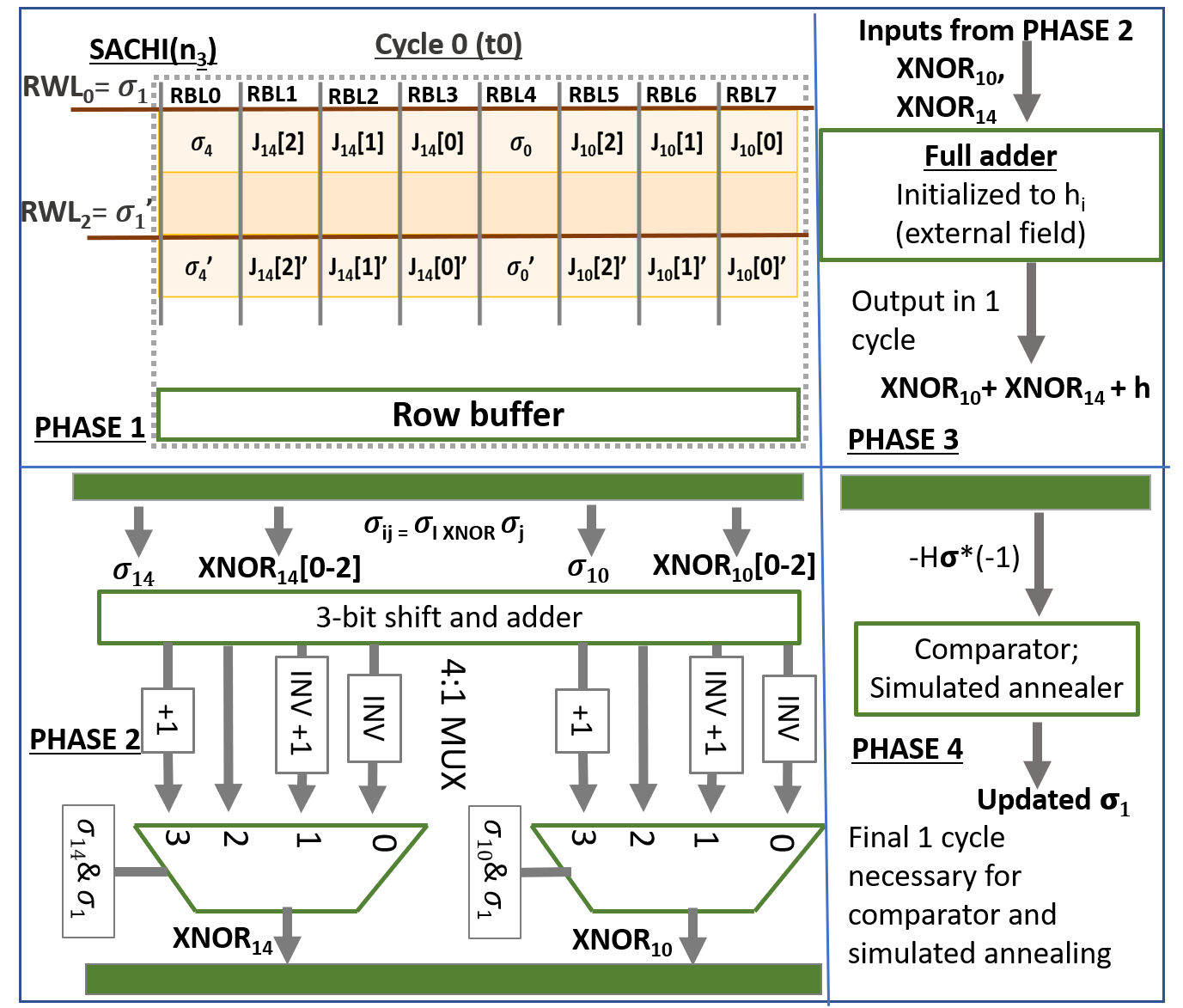}
\vspace{-2em}
\caption{\textbf{\underline{SACHI(n\textsubscript{3}):  Mixed stationary design} making use of reuse-aware compute to perform XNOR across all J\textsubscript{ij} bits and neighbors of target spin in a single cycle in Phase 1, improving maximum reuse to (neighbors)*(J\textsubscript{ij} resolution) (entire memory array highlighted). The decision logic and adder in Phase 3 are modified to support high throughput}}
\label{fig:ISCA(n3)}
\vspace{-0.25em}
\end{figure}
\subsubsection{\textbf{\underline{SACHI(n\textsubscript{3}) = Mixed stationary design}}}
 Although SACHI(n\textsubscript{2}) offers improved reuse and performance, the unique mapping between 2 spins for a particular IC and the way H\textsubscript{$\sigma$} is computed using eqn.\ref{eq:h_sigma} limits the reuse to R. 

    To further increase reuse, we propose a reuse-aware mixed stationary strategy, based on a few observations. (i) The spins are always binary-valued. (ii) The dot product between J\textsubscript{ij} and $\sigma$\textsubscript{j} is resolved as the dot product between J\textsubscript{ij} and $\sigma$\textsubscript{i}, if $\sigma$\textsubscript{j} and $\sigma$\textsubscript{i} are the same. If the spins are the same, the reuse-aware equation resolves to eqn.\ref{eq:h_sigma}, and inversion of XNOR output if the spins differ. The binary nature of spins makes equality checking of $\sigma$\textsubscript{j} and $\sigma$\textsubscript{i} an XNOR operation, which can be computed in parallel with XNOR of J\textsubscript{ij} and $\sigma$\textsubscript{i}.
This can be summarized as:
\begin{equation}\label{eq:reuse-aware}
    J\textsubscript{ij}*\sigma\textsubscript{j}=
    \begin{cases}
        J\textsubscript{ij}  $XNOR$ \sigma\textsubscript{i}, & \sigma\textsubscript{i} > 0 \;\& \; \sigma\textsubscript{i} $XNOR$ \sigma\textsubscript{j}=1  \\
        J\textsubscript{ij} $XNOR$ \sigma\textsubscript{i} + 1, & \sigma\textsubscript{i} < 0 \;\& \;
        \sigma\textsubscript{i} $XNOR$ \sigma\textsubscript{j}=1 \\
        J\textsubscript{ij} $XOR$ \sigma\textsubscript{i}, & \sigma\textsubscript{i} > 0 \;\& \;
        \sigma\textsubscript{i} $XNOR$ \sigma\textsubscript{j}=0 \\
         J\textsubscript{ij} $XOR$ \sigma\textsubscript{i} + 1, & \sigma\textsubscript{i} < 0 \;\& \;
        \sigma\textsubscript{i} $XNOR$ \sigma\textsubscript{j}=0 
        \\
    \end{cases}
\end{equation}
{\underline {$\sigma$\textsubscript{i} is mapped onto RWL;  J\textsubscript{ij} and $\sigma$\textsubscript{j} are stored in the compute}} 
{\underline {array, making it a mixed stationary design.}} $\sigma$\textsubscript{i} is shared across a complete row with no requirement of bitline select, reading all columns, decreasing the column circuitry's complexity. 
Consider a 4-spin network with a target spin ($\sigma$\textsubscript{1}) connected to spins $\sigma$\textsubscript{2-4}. ICs are stored in a row within the compute array, and the required computes are J\textsubscript{12}*$\sigma$\textsubscript{2}, J\textsubscript{13}*$\sigma$\textsubscript{3}, and J\textsubscript{14}*$\sigma$\textsubscript{4}. From a memory design perspective, if we store J\textsubscript{12}, J\textsubscript{13}, and J\textsubscript{14} in a row, we can perform J\textsubscript{12}*$\sigma$\textsubscript{i},  J\textsubscript{13}*$\sigma$\textsubscript{i} ,  J\textsubscript{14}*$\sigma$\textsubscript{i},  as all elements in a row share the same WL (mapped to $\sigma$\textsubscript{i}).  In the reuse-aware architecture, we map $\sigma$\textsubscript{i} to be $\sigma$\textsubscript{1}, enabling us to compute all dot product operations concurrently.  However, it's crucial to verify whether $\sigma$\textsubscript{1}($\sigma$\textsubscript{i}) is the same as $\sigma$\textsubscript{2}, $\sigma$\textsubscript{3}, $\sigma$\textsubscript{4} to get the correct H\textsubscript{$\sigma$} value.
\par The compute as illustrated in Fig.~\ref{fig:ISCA(n3)} proceeds as follows: In \textbf{phase 1}, the throughput of the SRAM array is as high as (N*R), as all XNOR computations of $\sigma$\textsubscript{i} are done in parallel. This high throughput enables compute without the usage of XNOR queue. \textbf{Phase 2} performs shift and add logic for computing the XNOR dot products followed by the decision logic to select between the 4 options in eq.\ref{eq:reuse-aware} to compute the reuse aware equation of H\textsubscript{$\sigma$}. 
    In \textbf{phase 3}, adder 
    performs the accumulation of partial dot products at once, to support the high throughput. 

    \par The time to compute H\textsubscript{$\sigma$} is independent of R and N, implying O(1) compute and the target spin is used across the entire row, satisfying IC criterion with reuse of N*R. 



\par \subsection{{Software support}} 
\begin{figure}[t]
\centering
\includegraphics[width=\linewidth]{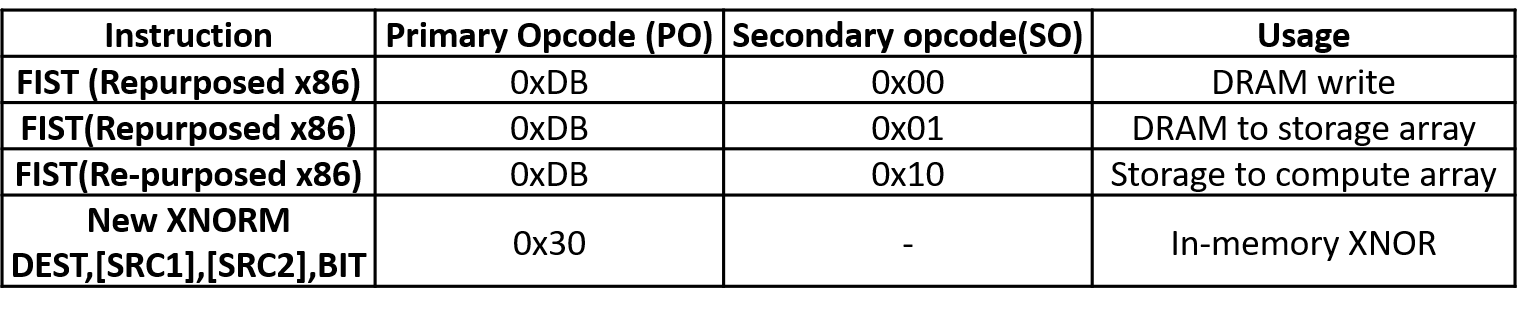}
\vspace{-2em}
\caption{\textbf{\ul{Software support} - FIST in x86 with different secondary opcodes to indicate movement from DRAM to storage/compute array. New XNORM instruction with SRC1 = the storage array address that is mapped onto RWL, SRC2 = the compute array address, BIT = J\textsubscript{ij} resolution and DEST = register for storing XNOR}}
\label{Software_support}
\vspace{-1.5em}
\end{figure}

The overhead in terms of additional compiler support is minimal for SACHI, as CPU assembly instructions can be repurposed to support SACHI because of the repurposable nature of SACHI. For instance, FIST (integer store), in x86 64-bit ISA, has a primary opcode (PO) of 0xDB without using a secondary opcode (SO). SO is used to indicate SACHI requests. A secondary opcode of 0x00 refers to DRAM write, 0x01 for DRAM to storage array write, 0x10 for transfer from storage to compute array. XNORM instruction (XNORM DEST,[SRC1],[SRC2], BIT), where SRC1= the address mapped onto RWL, SRC2= address of compute array, BIT = J\textsubscript{ij} resolution, is added to perform PIM XNOR, followed by near-memory reuse-aware compute (Fig.\ref{Software_support}).

\section{EXPERIMENTAL METHODOLOGY}
\subsubsection{\textbf{\underline{Configuration}}} SACHI configuration incorporating (i) a compute array consisting of 16 tiles, each tile (size 10KB) capable of storing 100 spins and 8-bit ICs, (ii) a storage array (size 160KB)  with 2 read ports (iii) digital peripheral logic sufficient to support compute array throughput is used for the experiments. The power/performance (PP) estimation takes into account data movement from the storage-compute array, in-memory compute and digital logic. 

\par \subsubsection{\textbf{\underline{Benchmarks}}} 
Experiments use the following benchmarks, with the size of variables from 500-1M.  We use the Ising formulation in \cite{Ising_formulation} to model these benchmarks. In all these benchmarks, we formulate the problem as H\textsubscript{$\sigma$} = -$\Sigma$J\textsubscript{ij}*$\sigma$\textsubscript{j} or H = -$\Sigma$J\textsubscript{ij}*$\sigma$\textsubscript{i}*$\sigma$\textsubscript{j}. This enables compute in all Ising machines, that can support dot-product of J\textsubscript{ij} and $\sigma$\textsubscript{j}. Therefore, no benchmark-dependent configuration of Ising machines is required, as long as the Ising machines are generic enough to compute the underlying graphs. These machines compute H\textsubscript{$\sigma$} for all workloads, assuming a certain type of graph and initialized spins/ICs associated with the different workloads.
\par  \textbf{a) Asset allocation:} Given m assets with \$80M value, divide the assets (J\textsubscript{ij} represents value) equally between 2 people. This is provided as an example of number partitioning in \cite{Ising_formulation}. This problem can be formulated as checking if H\textsubscript{$\sigma$} = $\Sigma$J\textsubscript{ij}*$\sigma$\textsubscript{j} is zero. Here, $\sigma$\textsubscript{j}=+1/-1 helps distinguish between the 2 people, J\textsubscript{ij} signifies the value of each asset.  This is a sparsely connected graph of number of spins equal to m. 
\par \textbf{b) Image segmentation:} Given a densely connected image of m*n pixels, this benchmark identifies the max cut that splits the image into foreground and background (spin +1/-1), with J\textsubscript{ij} (edge weight) indicating the difference in pixel values between neighbors. H\textsubscript{$\sigma$} formulation is the same as that of formulation given for max cut in \cite{BRIM}\cite{Ising_formulation}
\par \textbf{c) Traveling salesman:} Given a network of connected cities (spins), shown as a complete graph, this problem finds a route with total distance lesser than W between 2 given cities. The decision version of traveling salesman mentioned in \cite{Ising_formulation} is used. This problem checks if H = $\Sigma$ J\textsubscript{ij} * $\sigma$\textsubscript{i} * $\sigma$\textsubscript{j} $<$ W between the 2 cities. J\textsubscript{ij} is the distance, $\sigma$\textsubscript{i} \& $\sigma$\textsubscript{j} = 1, implies a route between the two cities. This is achieved by solving H\textsubscript{$\sigma$} = -$\Sigma$ J\textsubscript{ij} * $\sigma$\textsubscript{j} and checking if the associated H is lesser than W. 
\par \textbf{d) Molecular dynamics:} Given a set of atoms in a molecule connected as King's graph, this identifies the atomic spin states in the lowest energy configuration.
This is to show readers that J\textsubscript{ij} in ferro-magnetism in eqn.1 is the force of attraction between 2 neighboring atoms \cite{isinghistory}.

\subsubsection{\textbf{\underline{SRAM/Data movement Power/Perf (PP) estimation}}} SRAM data array with peripheral circuits are designed using Cadence Virtuoso and the device parameters are modeled using FreePDK 45nm technology\cite{45nmPDK}. In-memory XNOR compute energy is measured when RWL is turned ON, and RBL is discharged. To measure RWL energy (pJ/bit), RWL under-driven approach \cite{SST_phase}\cite{fefet} with RWL capacitance of 50fF is used. To calculate the discharge energy of RBL (pJ/bit), 35fF RBL capacitance\cite{SST_phase} is assumed for the SRAM array of 100 rows/columns (to match the size of a compute tile) for SRAM operating voltage of 1V. 
The compute latency of the SRAM array is found to be 2ns. The data movement energy for mapping RWL onto compute array is 1pJ/bit, assuming that the data movement energy is $\sim$800x the addition energy \cite{Mutlu}, with 100ns latency for storage to compute array movement. Furthermore, for applications whose storage requirement is greater than compute array size, (i) energy costs include data movement from DRAM to the storage array, SRAM write, and DRAM controller logic for prefetch (ii) performance cost includes SRAM write latency. 

 \subsubsection{\textbf{ \underline{Digital logic PP estimation}}} Synthesis of digital circuits using Synopsis RTL Design Compiler is used to quantify power/energy having a cycle time of 5ns for 45nm technology node and operating voltage of 1V.  We have assumed a clock cycle time of 5ns solely because of the slower standard-cell logic gates in 45nm technology node used for simulations. 45nm PDK was chosen because it is open-sourced. With performance scaling, as (technology node) as mentioned in \cite{Conv_SRAM}, the cycle time of SACHI can match modern CPUs. 

 \subsubsection{\textbf { \underline{BRIM/Ising-CIM Comparison}}}  To understand trade-offs between proposed and existing designs, factors include \textbf{(a)} storing input variables and ICs onto DRAM (necessary for BRIM/Ising-CIM/SACHI), \textbf{(b)} loading variables onto storage/compute arrays from DRAM (necessary for Ising-CIM/SACHI), data movement from DRAM to coupled oscillator (necessary for BRIM) and \textbf{(c)}Ising compute. Across all designs, loading (both into DRAM(a) and from DRAM(b)) involves data movement at 64B per cycle, and the number of cycles depends on COP size. For example, a COP with 100 spins King’s Graph, 8-bit J\textsubscript{ij}, requires $\sim$13 cycles for storage onto DRAM, with a fixed loading energy cost of 1pJ/bit. 
 The technology used is 45nm, cycle time of 5ns, and digital logic synthesis in 45nm to ensure a fair comparison across designs. \textbf{BRIM} evaluation considers Ising compute using a coupled oscillator, and associated digital logic/DAC. H compute in coupled oscillator+DAC takes 4-13 cycles per iteration. In the best case (used for comparison with SACHI), 1 cycle is needed for each of memory array read, DAC, and compute using coupling oscillator topology and annealing control. However, considering additional cycles for precharge, write into the memory array, along with a sequential DAC, a total of 13 cycles are needed for H compute. Typically, physical Ising machines can compute multiple spins in parallel. However, in BRIM, this is limited by 2 factors: i) The presence of storage capacitor introduces delays in capturing voltage changes, restricting spin initially at '0' to undergo fast transition from '0' to '1'. This is especially when input voltage transitions from its neighboring spins are abrupt, which is most likely the case. ii) Leakage through unconnected paths (spins that are unconnected to each other-similar to an unconnected cell in the case of DRAM), from passive capacitor is especially important when the node voltage is close to the trip-point of the ZIV diode. The power for coupled oscillator logic is 250mW for 2000 spins (100 neighbors per spin) and is proportional to the number of spins and neighbors. For a single 8-bit DAC, the power is $\sim$0.004mW, and there are 16 banks (1 DAC per bank) with associated digital logic consisting of 16:1 8-bit multiplexers (to map DAC outputs onto coupling units)/(16*8) flops per bank (storing the output of DAC). In contrast, SACHI has no DAC/associated digital logic, saving on performance and power.
\textbf{Ising-CIM} compute involves XNOR compute in eDRAM, and annealing control. XNOR compute requires 3 cycles each for computing the updated spin values and performing the update, scaling with the number of spins and neighbors due to lack of parallelism. XNOR in eDRAM requires 1.2x power compared to 8T SRAM due to increased operating voltage. Annealing power is the same for all designs.

\section{RESULTS}

\par \subsubsection{\textbf{\underline{Comparison with BRIM}}}  
\begin{figure}[t]
\centering
\includegraphics[width=\linewidth]{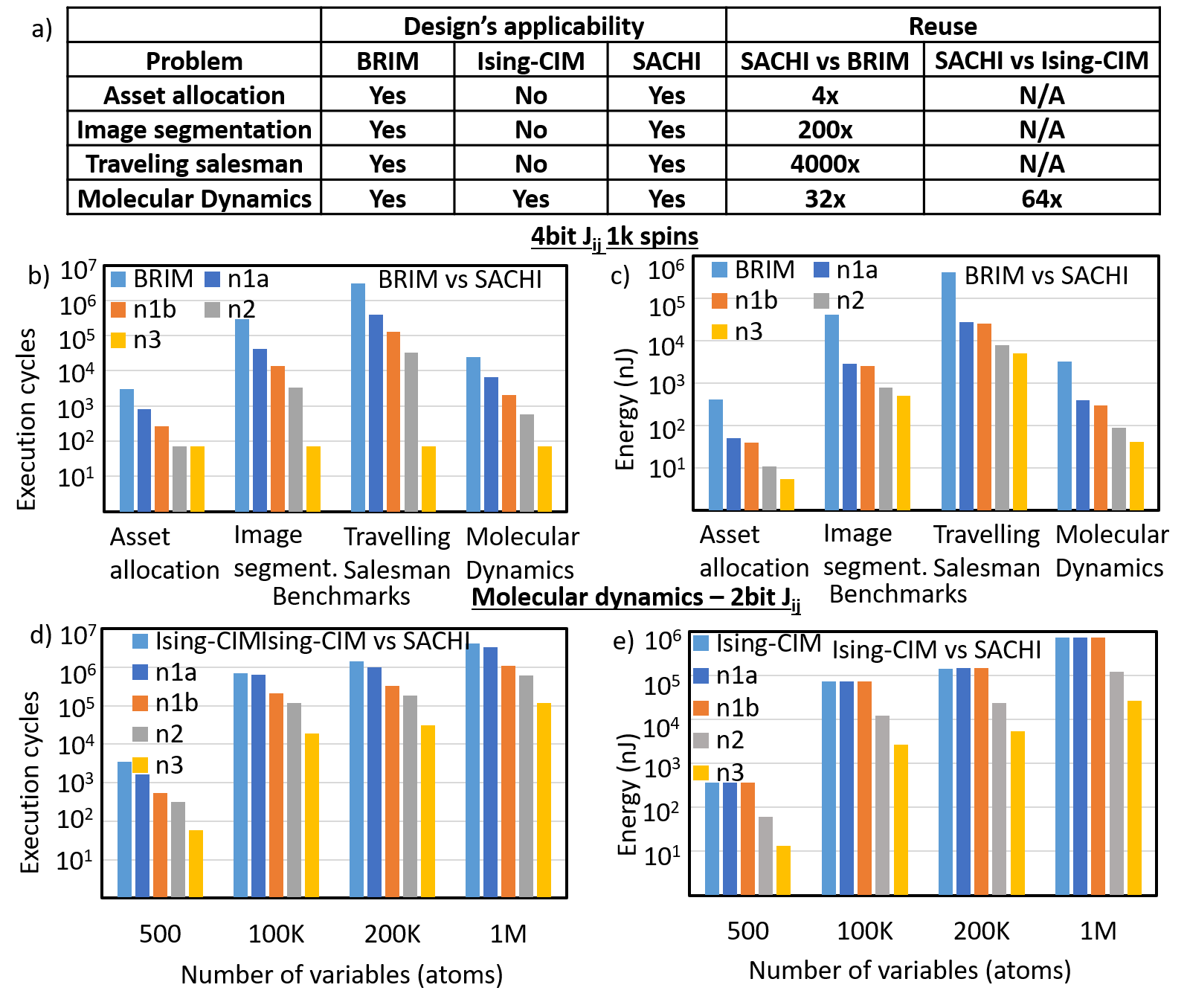}
\vspace{-2em}
\caption{\textbf{SACHI comparison with BRIM and Ising-CIM for reuse and designs applicability to different COPs. \underline{SACHI comparison with BRIM -} b) Number of cycles  c) total energy to solve COP including loading \textbf{\underline{SACHI comparison with Ising-CIM}} - e) Number of cycles  f) total energy to solve COP including loading}}
\label{BRIM_comparison}
\vspace{-1em}
\end{figure}

BRIM is compared with SACHI using all benchmarks, assuming 1K spins and 4-bit interaction coefficients in (Fig.\ref{BRIM_comparison}).

\par \textbf{a) Performance:} SACHI(n\textsubscript{3}) performs $\sim$36x better, for asset allocation as shown in Fig.\ref{BRIM_comparison}b. For large COPs like traveling salesman with high graph connectivity,  SACHI(n\textsubscript{3}) performs $\sim$300x better including the loading effect, as SACHI(n\textsubscript{3}) offers high parallelism across neighbors per node, while BRIM does not. In benchmarks with lesser graph connectivity and loading overhead, like image segmentation and molecular dynamics, SACHI(n\textsubscript{3}) performs $\sim$286x, $\sim$160x better as the opportunity for parallelism across neighbors is less. 
\par \textbf{b) Energy:} 
SACHI(n\textsubscript{3}) improves energy by $\sim$72x for asset allocation, $\sim$80x for image segmentation, and $\sim$79x for molecular dynamics as shown in Fig.\ref{BRIM_comparison}c. The need for greater data transfer for traveling salesman results in a slight reduction in energy efficiency (although still has 75x improvement). This is because the Ising graph's fully connected nature necessitates retrieval of more spins/ICs from DRAM/L2 into L1 cache. This increased data transfer incurs additional power consumption leading to a greater overall energy demand. The increased data movement requirement in traveling salesman leads to slightly degraded energy, but still $\sim$75x energy improvement (Fig.\ref{BRIM_comparison}c).
\par 
The major reason for SACHI's superior performance is its increased reuse. In BRIM, reuse is 1, as one IC fetched from memory is used in 1 H compute in the coupling units in BRIM. The reuse in SACHI(n\textsubscript{3}) is $\sim$200x for image segmentation, $\sim$4000x for travelling salesman, $\sim$32x for molecular dynamics (tabulated in Fig.\ref{BRIM_comparison}a). 

 \subsubsection{\textbf{\underline{Comparison with Ising-CIM}}}  Ising-CIM is limited to King's Graph with 2-bit unsigned ICs. Hence we restrict our comparison to 2-bit molecular dynamics COP. %
\par \textbf{a) Performance}:  SACHI(n\textsubscript{3}) performs $\sim$70x/$\sim$80x better for 500/1M atoms (including loading) (Fig.\ref{BRIM_comparison}d). This is because of the parallelism across neighbors and J\textsubscript{ij} resolution in SACHI(n\textsubscript{3}), while Ising-CIM does not have such parallelism. 

\par \textbf{b) Energy}: SACHI(n\textsubscript{1a})/SACHI(n\textsubscript{1b}) have the same energy because the increased performance in SACHI(n\textsubscript{1b}) is compensated by decreased power in SACHI(n\textsubscript{1a}). The energy improvement in SACHI(n\textsubscript{3}) for 500/1M spins is $\sim$40x/$\sim$75x respectively (Fig.\ref{BRIM_comparison}e) because of the increased voltage requirement for eDRAM and reduced performance in Ising-CIM.
\par 
The maximum reuse is 1 in Ising-CIM, as every bit of IC is used only in 1 H\textsubscript{$\sigma$} compute performed in the eDRAM compute array. This leads to SACHI(n\textsubscript{3}) providing $\sim$16x more reuse in molecular dynamics (Fig.\ref{BRIM_comparison}a tabulates the reuse factors).

\par \subsubsection{\textbf{\underline{{Comparison with Other Optimized Solvers}}}} 
\begin{figure}[t]
\centering
\includegraphics[width=\linewidth]{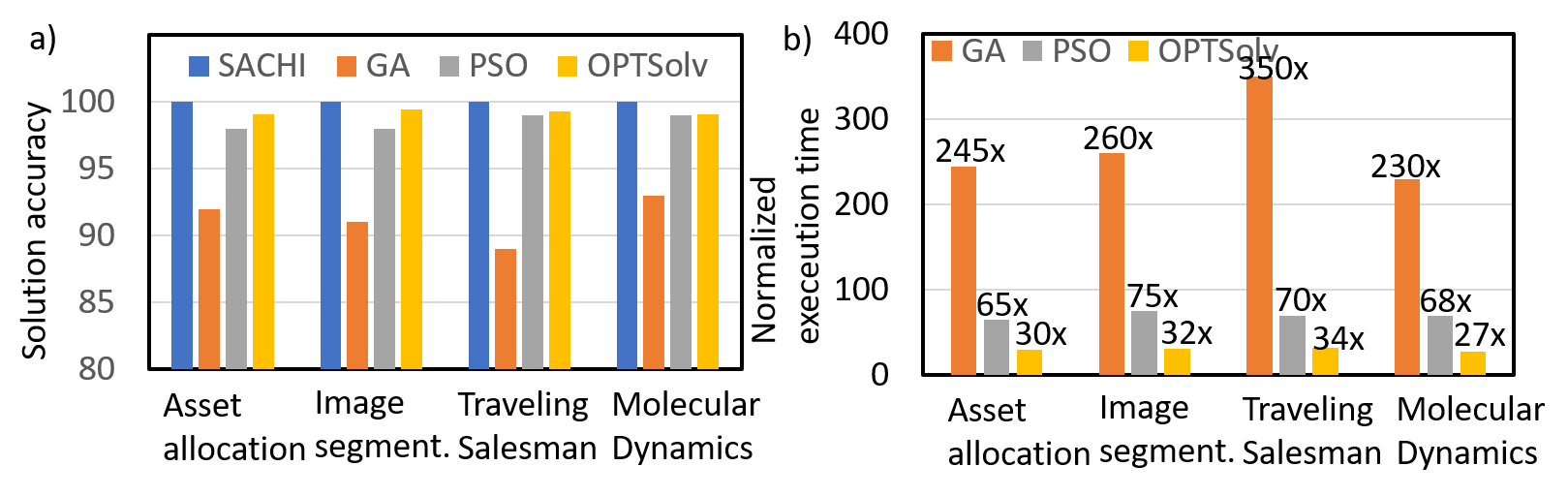}
\vspace{-2em}
  \caption{\textbf{Comparison with GA/PSO/optimized solvers (OPTSolv) - Solution accuracy, normalized execution time wrt SACHI for asset allocation, image segmentation, traveling salesman and molecular dynamics}}
\label{GA_comparison}
\vspace{-1em}
\end{figure}

SACHI's performance in comparison to solutions implemented using Genetic Algorithm (GA), Particle Swarm Optimization (PSO), and dedicated Optimized Solvers is of interest. GA/PSO is implemented on an Intel i5-8265U CPU running at 3.9GHz. 
For the GA implementation, GAlib \cite{GAlib} is used.
The solution quality is measured by comparing GA's accuracy to SACHI's 100\% accuracy. GA's accuracy is lower, as shown in (Fig.\ref{GA_comparison}a). This could be attributed to GA's global-only search for selecting the best candidates in each generation. In contrast, Ising/PSO performs updates based on neighbors, resulting in faster convergence (Fig.\ref{GA_comparison}b). In PSO, the selection criterion considers personal best (pbest) and global best (gbest) for all candidates, where pbest is compared against gbest at the end of each iteration to update the fitness. 
{\underline {The comparison with dedicated optimized solvers/algorithms} }{\underline{(OPTSolv)}} for benchmarks such as Concorde solver for traveling salesman, Ford-Fulkerson network flow for image segmentation, LAMMPS \cite{LAMMPS} for molecular dynamics, is also presented in Fig.\ref{GA_comparison}. 
SACHI outperforms these solvers by 27-34x because of parallelism across N and R from reuse-aware near-memory compute. 

\subsubsection{\textbf{\underline{Scalability}}}  
\begin{figure}[b]
\centering
\includegraphics[width=\linewidth]{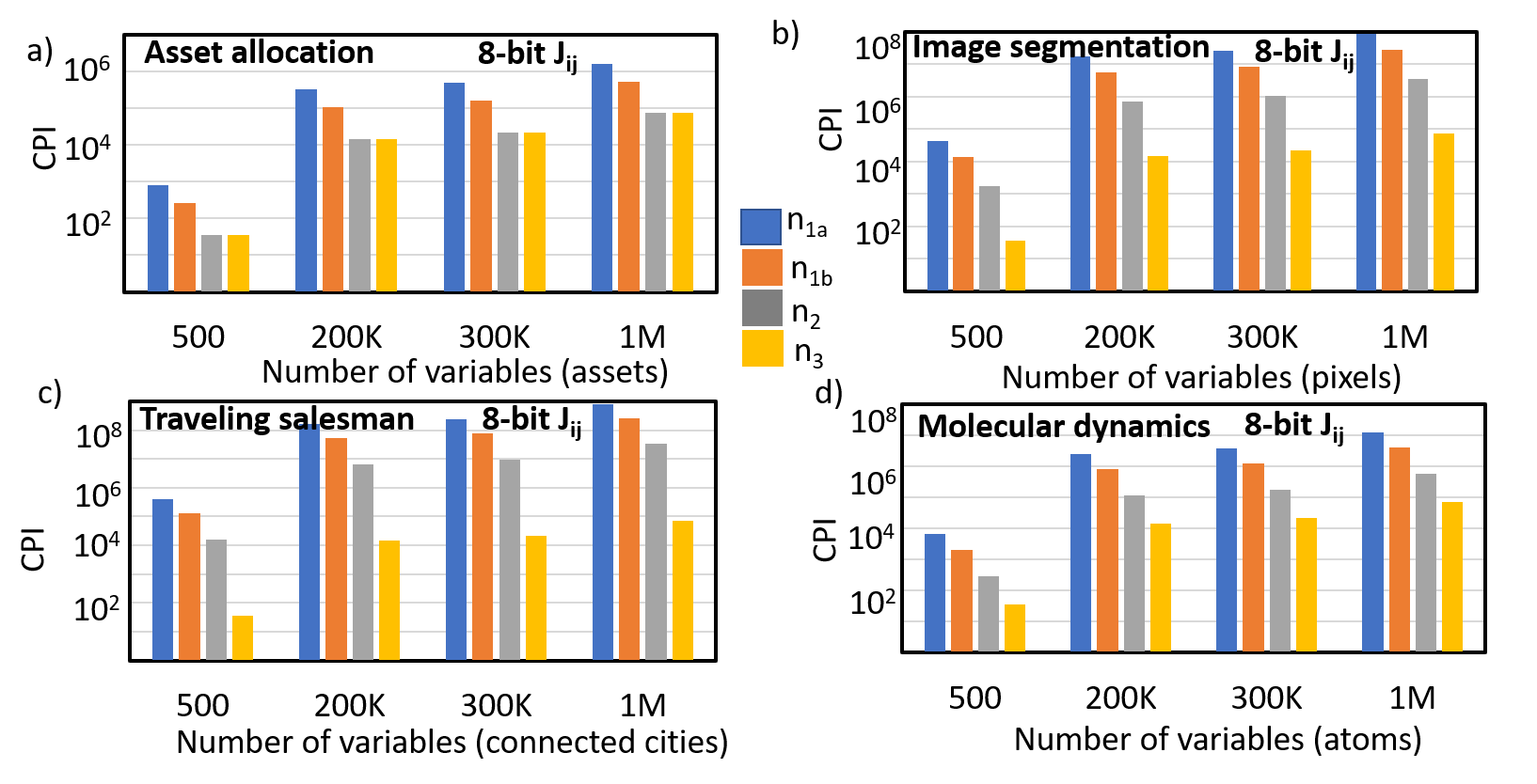}
\vspace{-2em}
\caption{\textbf{\underline{Scalability} - Cycles Per Hamiltonian iteration (I) with increasing variable size for a) Asset allocation b) Image segmentation c) Traveling salesman and d) Molecular dynamics COPs showing SACHI's scalability to different large variable count COPs}}
\label{Scalability}
\vspace{-0.25em}
\end{figure}

SACHI is highly scalable for graphs with diverse connectivity, accommodating various sizes without limitations. Fig.\ref{Scalability} illustrates its performance as CPI (clock cycles per Hamiltonian iteration) for spin counts ranging from 500 to 1M, showcasing the impact of compute array overflow in each scenario. For instance, in the case of (a) 500 spins: Spins fit inside the compute array for all SACHI designs, (b) 200K spins: Spins fit inside the compute array for all SACHI designs except SACHI(n\textsubscript{3}), (c) 300K spins: Spins do not fit inside the compute array for SACHI(n\textsubscript{2}) and SACHI(n\textsubscript{3}), (d) 1M spins: Spins do not fit inside the compute array for all SACHI designs. A few noteworthy points are as follows: (i) SACHI(n\textsubscript{3}) demonstrates the highest performance across all workloads due to the parallel compute of all spin neighbors and IC bits. (ii) SACHI(n\textsubscript{2}) and SACHI(n\textsubscript{3}) share the same CPI (as there is 1 neighbor/spin), except for 200K spins in asset allocation where SACHI(n\textsubscript{2}) outperforms SACHI(n\textsubscript{3}) due to SRAM write latency, as SACHI(n\textsubscript{3}) has a filled compute array and needs rewriting for the next round of compute. (iii) SACHI(n\textsubscript{1a}) shows moderate performance due to inefficient load balancing among compute tiles with adjacent spins in the same tile. SACHI(n\textsubscript{1b}) resolves this issue by re-organizing neighboring spins across different tiles. (iv) Traveling salesman benchmark has the highest CPI, primarily due to the high connectivity of the underlying complete graph, affecting SACHI(n\textsubscript{1}) and SACHI(n\textsubscript{2}) designs, whose performance depends on the number of neighbors per spin. (v) For many COPs, 1M spins are large enough, but for image segmentation, 2M pixels (HD video) and 8M pixels (UHD video) cases were studied. They take 10\textsuperscript{9} and 2*10\textsuperscript{10} CPI, respectively.

\subsubsection{\textbf{\underline{Reconfigurability}}} 
\begin{figure}[t]
\centering
\includegraphics[width=\linewidth]{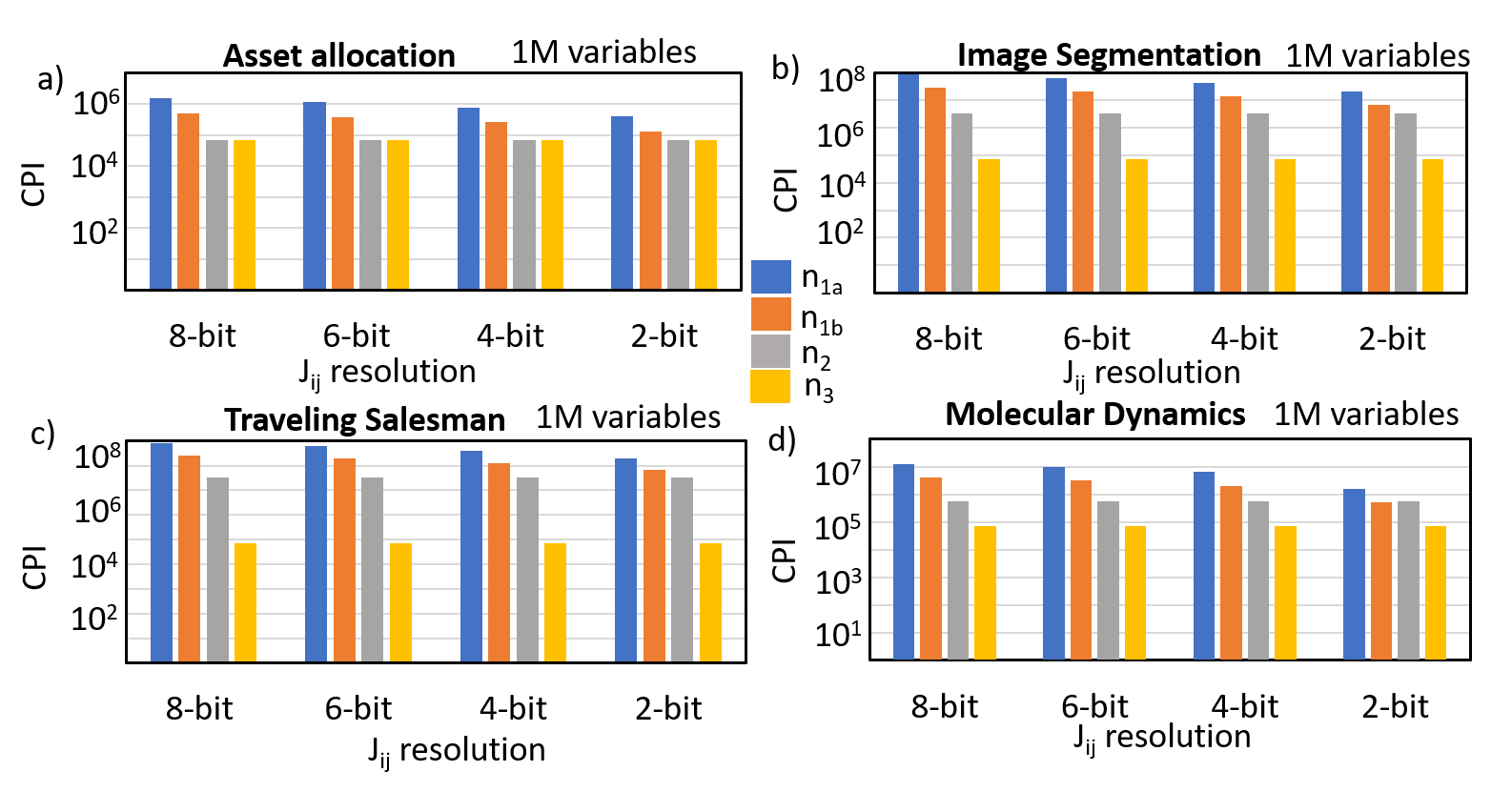}
\vspace{-2em}
\caption{\textbf{\underline{Reconfigurability} - Cycles Per Iteration (CPI, i.e. Hamiltonian iterations) with decreasing IC resolution: a) Asset allocation b) Image segmentation c) Traveling salesman d) Molecular dynamics showing the feasibility of SACHI being reconfigurable to different IC resolution}}
\label{Reconfigurability}
\vspace{-1em}
\end{figure}

Reconfigurability is shown by using SACHI for different J\textsubscript{ij} resolutions, configured based on COP. Fig.\ref{Reconfigurability}a-d shows the sensitivity of CPI for 1M spins and J\textsubscript{ij} resolution from 2 to 8 bits. SACHI(n\textsubscript{2}) and SACHI(n\textsubscript{3}) show no change in CPI, as the performance is independent of J\textsubscript{ij} resolution. SACHI(n\textsubscript{1a}), SACHI(n\textsubscript{1b}) show performance improvement with reduced resolution, as there are fewer in-memory XNOR computes.

\par \subsubsection{\textbf{\underline{Time to solution/Solution quality}}} 
\begin{figure}[t]
\centering
\includegraphics[width=\linewidth]{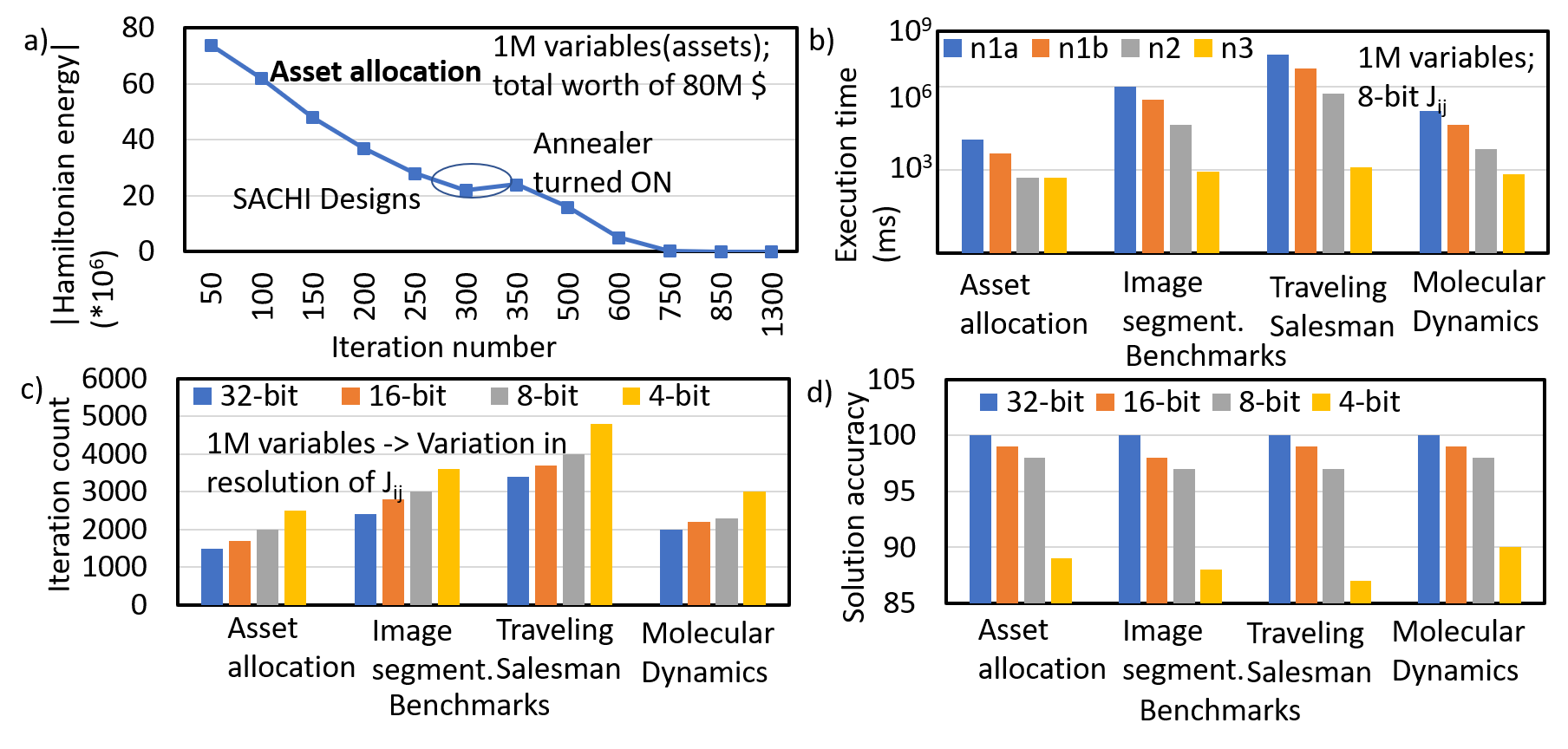}
\vspace{-2.5em}
\caption{\textbf{\underline{Time to solution/solution accuracy} - a) Hamiltonian energy (H) reduction with increasing iteration count for Asset allocation, using simulated annealing (SA) to move out of local minima. b) Solution time improvement from SACHI(n\textsubscript{1}) to SACHI(n\textsubscript{3}) architecture due to reuse aware compute c) Increased iteration count with lower resolution for IC due to frequent requirement of SA steps d) Solution accuracy degradation with lower IC resolution under iso-performance condition}}
\label{Time_solution}
\vspace{-1em}
\end{figure}

The time to solution depends on the number of iterations needed to converge at a solution, CPI, and the clock period. The solution convergence is reached when the Hamiltonian energy (H) remains unchanged, and simulated annealing cannot further reduce the energy. 
Fig. \ref{Time_solution}a) illustrates the H change with iteration number for 1M assets in the asset allocation COP. Simulated annealing's contribution to the overall execution time is less than 1\%, enhancing accuracy by 0.8\%. It is implemented by probabilistically flipping based on the Metropolis acceptance criterion \cite{Shanshan_Ising}, comparing likelihood against a predefined value within the annealer block (Sec.4).  The number of iterations across SACHI designs is the same, as they all arrive at the same H at the end of each iteration. The execution/annealing time is the highest for traveling salesman COP because of the high degree of graph connectivity. Asset allocation requires the fewest iterations, followed by molecular dynamics, owing to limited graph connectivity (Fig.\ref{Time_solution}b). To assess the accuracy loss due to resolution reduction, the change in the number of iterations needed to converge at a solution with J\textsubscript{ij} resolution is observed. An 8-bit J\textsubscript{ij} resolution strikes the best balance between reduced memory footprint requirement and loss in accuracy. Reducing the resolution below 8-bit (e.g., 4-bit) leads to a sharp increase in the number of iterations (Fig.\ref{Time_solution}c). For 32-bit resolution, the number of iterations is the lowest, as accurate compute decreases the chances of being trapped in locally optimal solutions and reduces the need for simulated annealing iterations.Fig.\ref{Time_solution}d)  shows the solution accuracy for lower J\textsubscript{ij} resolutions once the 32-bit J\textsubscript{ij} has converged at the solution. While 4-bit reduces accuracy below 90\%, 8-bit retains accuracy with a smaller memory footprint than 32-bit.

\section{Discussion}
 \par \subsubsection{\textbf{\underline{Impact on conventional workloads}}} SACHI does not detrimentally affect other workloads. 
 The L1 datapath is split as (a) fill into L1 from L2 and (b) read from L1. The fill datapath does not get impacted at all. The read from L1 cache is not affected because (i) we do not make modifications to the memory array for PIM (L1 caches in modern processors are already 8T) (ii) although there is an additional 2:1 multiplexer delay in the L1 controller/periphery to choose between normal and compute mode of L1 cache, that can be easily retimed and absorbed by synthesis tools. Furthermore, the additional near-memory peripheral logic does not intervene in normal operation as this is a separate datapath. 

 \par \subsubsection{\textbf{\underline{Impact of increased L1/L2 cache size}}} In our simulations, we have assumed L1/L2 cache sizes of 10KB/160KB because SRAM bit-cell parameters are designed in Virtuoso for the same memory capacity.  However, in modern CPUs, the typical L1 cache ranges from 64KB to 256KB, while L2 cache spans from 1MB to 8MB. By increasing the cache size, SACHI's performance can be further enhanced due to several factors: (i) increased parallelism across neighbors (N) and resolution (R) can be achieved (ii) higher resolution for ICs improves solution time and accuracy. (iii) the larger L1 cache can better accommodate large-sized COPs such as the traveling salesman problem. 
 Experiments show that for the traveling salesman problem with 1M spins, a 64KB/1MB cache offers 5x/8x better performance/energy compared to the 10KB/160KB configuration. These improvements lead to 16x performance/20x energy improvement for the 256KB/8MB cache with 1M spins. It is noteworthy that performance does not degrade for any benchmark. Though increasing cache size leads to a slight rise in power due to larger RBL/RWL capacitance in larger arrays, this is outweighed by the higher performance, resulting in overall energy gain.

\par \subsubsection{\textbf{\underline{Additional software details}}} We assume that the cache operates in a single mode at a time, either compute/normal mode. 
The mode switch can be achieved by programming a special-purpose register. Additionally, ongoing work includes (i) developing a CUDA-like library/API to program SACHI as part of a complete program, (ii) extending the library to support Ising formulation of COPs, similar to the OpenQASM framework \cite{openqasm}, and (iii) enabling the interruption of SACHI, storing contexts, ASIDs, and minimal payload in TLBs to facilitate rapid page translation during a context switch between modes (but these are outside the scope of this paper).

\section{Conclusion}
We presented SACHI, an iterative-compute based all-digital Ising architecture that employs reuse-aware stationarity schemes for Ising spins and interaction coefficients. It combines elements of re-purposability, scalability, reconfigurability, and reuse-aware data-stationary near-memory compute to achieve 160x/90x better performance, and improved energy of 79x/75x with better reuse of 32x/16x for molecular dynamics COP over BRIM/Ising-CIM, and 27x-34x over dedicated optimized solvers for different COPs.

\bibliographystyle{IEEEtranS}
\bibliography{refs}

\end{document}